\documentclass[%
 reprint,
 amsmath,amssymb,
 aps,
nofootinbib
]{revtex4-2}

\usepackage{graphicx}
\usepackage{dcolumn}
\usepackage{bm}
\usepackage{hyperref}
\usepackage{enumitem}
\usepackage{times}
\usepackage[normalem]{ulem}

\usepackage{xcolor}

\begin{document}


\title{Cosmological multifield emulator}

\author{Sambatra Andrianomena$^{1,2}$}\email{hagatiana.andrianomena@gmail.com}
\author{Sultan Hassan$^{2,3}$}\email{sh7530@nyu.edu}
\author{Francisco Villaescusa-Navarro$^{4}$}
\affiliation{%
 $^{1}$SARAO, Liesbeek House, River Park Liesbeek Parkway, Settlers Way, Mowbray, Cape Town, 7705}
 \affiliation{%
 $^{2}$Department of Physics \& Astronomy, University of the Western Cape, Bellville, Cape Town 7535,
South Africa}
  \affiliation{%
 $^{3}$Center for Cosmology and Particle Physics, Department of Physics, New York University, 726 Broadway, New York, NY 10003, USA}
 \affiliation{%
 $^{4}$ Center for Computational Astrophysics, Flatiron Institute, 162 5th Ave, New York, NY 10010, USA}




\date{\today}

\begin{abstract}
 We present the application of deep networks to learn the distribution of multiple large-scale fields, conditioned exclusively on cosmology while marginalizing over astrophysics. Our approach develops a generalized multi-field emulator based purely on theoretical predictions from the state-of-the-art hydrodynamic simulations of the CAMELS project, without incorporating instrumental effects which limit the analysis to specifics of a particular large-scale survey design. 
To this end, we train a generative adversarial network to generate images composed of three different channels that represent gas density (Mgas), neutral hydrogen density (HI), and magnetic field amplitudes (B).   
We consider an unconstrained model and another scenario where the model is conditioned on the matter density $\Omega_{\rm m}$ and the amplitude of density fluctuations $\sigma_{8}$.
We find that the generated images exhibit great quality which is on a par with that of data, visually. Quantitatively, we find that our model generates maps whose statistical properties, quantified by probability distribution function of pixel values and auto-power spectra, agree reasonably well up to the second moment with those of the real maps. Moreover, the mean and standard deviation of the cross-correlations between fields in all maps produced by the emulator are in good agreement with those of the real images, which indicates that our model generates instances whose maps in all three channels describe the same physical region. 
Furthermore, a CNN regressor, which has been trained to extract $\Omega_{\rm m}$ and $\sigma_{8}$ from CAMELS multifield dataset, recovers the cosmology from the maps generated by our conditional model, achieving $R^{2}$ = 0.96 and 0.83 corresponding to $\Omega_{\rm m}$ and $\sigma_{8}$ respectively. This further demonstrates the great capability of the model to mimic CAMELS data. Our model can be useful for generating data that are required to analyze the information from upcoming multi-wavelength cosmological surveys.  
\end{abstract}

\maketitle



\section{\label{sec:Intro}INTRODUCTION}
Multiwavelength astronomy has been identified as a priority in the 2020 decadal survey~\citep{NAP26141}. Current and upcoming missions will survey the Universe at different wavelengths, from X-rays to radio: e.g. e-Rosita \citep{predehl2021erosita}, SKA \citep{dewdney2009square, weltman2020fundamental}, Euclid \citep{amiaux2012euclid}, DESI \citep{aghamousa2016desi}, CMB-S4 \citep{abazajian2019cmb}, Rubin \citep{ivezic2019lsst} and Roman \citep{mosby2020properties} Observatories, HIRAX \citep{newburgh2016hirax}, CHIME \citep{amiri2022overview}, Spherex \citep{dore2014cosmology}. The data from these missions is expected to help us improve our knowledge of both galaxy evolution and cosmology. In order to maximize the scientific outcome of these missions, accurate theoretical predictions are needed for different physical fields. Hydrodynamic simulations are the most sophisticated tools that can be used to study and model this. Their main drawback is their computational cost that, being very large, limits the volume and resolution of these simulations \citep{Vogelsberger_review}. Being able to speed up these simulations is critical in order to perform a large variety of tasks such as parameter inference and model selection. In addition, developing a fast emulator for large-scale fields to extract cosmological information, while marginalizing over those of astrophysics is an essential motivation behind this work. The main desire to marignalize over astrophysics is that different subgrid models of astrophysics (e.g. such as the recipes to form stars and AGNs as well as their driven feedback) are degenerate and tend to reproduce similar observations (e.g. see \cite{villaescusa2021camels, iyer2020diversity}).
Previous studies aimed at accelerating various cosmological simulations of different fields using generative models. The generative adversarial networks (GAN) model in \cite{rodriguez2018fast} was trained to speed up 2D cosmic web simulations. The GANs in \cite{mustafa2019cosmogan} learned  to reproduce high resolution ($1024\times1024$ pixels$^2$) weak lensing convergence maps. The Wasserstein Generative Adversarial Networks (WGANs), which were built in \cite{zamudio2019higan}, were capable of generating 3D HI maps of $64\times64\times64$ pixels at $z = 5$. \cite{perraudin2019cosmological} prescribed a scalable GAN method to emulate 3D $N$-body simulations cubes. Similarly, \cite{feder2020nonlinear} trained GANs to generate 3D cosmic web using dataset from GADGET-2 $N$-body simulations \citep{springel2005cosmological}. \cite{perraudin2021emulation} built a conditional GANs to learn 2D dark matter maps conditioned on $\Omega_{\rm m}$ and $\sigma_{8}$. \cite{curtis2022cosmic} investigated the void properties from maps of large scale structure which were generated by GANs. Other interesting approaches consist of mapping dark matter to other physical fields whose simulation can be expensive. Various studies used adversarial networks \cite{troster2019painting, harrington2022fast, andrianomena2023invertible} and other methods like variational autoencoder and normalizing flows \cite{horowitz2021hyphy, dai2021learning} to infer  physical quantities from dark matter.

In this work, based on the success of GANs\footnote{GANs and diffusion models remain the state-of-the-art techniques for image generation tasks on different benchmark datasets (see https://paperswithcode.com/task/image-generation). }, we make use of its variant to produce multifield images -- 2D maps whose each channel corresponds to a different physical field -- that have the desired statistical properties for each field individually but also their cross-correlations. This is crucial in order to extract most of the information from different large-scale surveys. We note that while the generation of multifield images has been performed in the past \cite[see e.g.][]{tamosiunas2021investigating} this work contributes to that direction by using a more sophisticated model and a richer and more complex dataset. More specifically, our aims are as follows:
\begin{enumerate}
    \item Develop a generalized theoretically-predicted multifield emulator within our scale of interest (25 Mpc/$h$)  without accounting for instrumental effects, which may restrict the analysis to a particular survey design. We focus on 25 Mpc/$h$ scales to probe non-linear fluctuations around the Baryon Acoustic Oscillations (BAO) scales and high frequency modes $k \sim  0.1-10\, h$/Mpc.
    \item Learn the conditional distribution of multifields solely on cosmological parameters to mariginalize over the poorly constrained astrophysical subgrid models - a primary goal of the CAMELS project \cite{villaescusa2021camels}.
\end{enumerate}
The paper is structured as follows: we provide a quick overview of generative adversarial networks and the model that is used in this work in \S~\ref{sec:gan}. The dataset considered in our analyses is discussed in \S~\ref{sec:data}, and the results are presented in \S~\ref{sec:results}. We finally conclude in \S~\ref{conclusions}.


\section{Generative adversarial networks}\label{sec:gan}
To build a generative model, \cite{goodfellow2014generative} prescribed a novel approach that leverages the concept of a zero-sum non-cooperative game, i.e. one wins to the detriment of the other. The method consists of two competing network models in which a generator $G$ approximates the distribution $p_{g}$ over the real data ${\bm x}$ by trying to fool a discriminator $D$ which learns to estimate the probability $D({\bm x})$ that a given example ${\bm x}_{i}$ is real. During training, the generator learns to map a prior noise distribution $p_{z}$ to data space $G({\bm z})$. The discriminator attempts to differentiate the real from the fake examples. The parameters of $D({\bm x})$ and $G({\bm z})$ are updated by maximising ${\rm log}(D({\bm x}))$ and minimising ${\rm log}(1 - D(G({\bm z})))$ respectively, similar to optimizing a value function $V(D,G)$ \textit{of a two-player min-max game} according to \citep{goodfellow2014generative}
\begin{eqnarray}\label{value-function}
&& \underset{G}{min}\ \underset{D}{max}\ V(D,G) = \mathbb{E}_{{\bm x}\sim p_{\rm{data}}({\bm x})}[{\rm log}(D({\bm x}))]\nonumber \\ 
&& ~~~~ + \mathbb{E}_{{\bm z}\sim p_{\bm z}({\bm z})}[{\rm log}(1-D(G({\bm z})))].
\end{eqnarray}
A solution is optimal when the generator is capable of producing samples $G({\bm z})$ with features similar to those of real data such that the discriminator $D$ acts like a classifier with a random guess when identifying real and fake images from the training data and generator respectively. In other words, as the training converges, the discriminator \citep{goodfellow2014generative} $$ D^{*}_{G} = \frac{p_{\rm data}({\bm x})}{p_{\rm data}({\bm x})+p_{g}({\bm x})}$$ tends to $D^{*}({\bm x})\sim \frac{1}{2}$ as $p_{g}$ approaches $p_{\rm data}$.
\begin{figure*}
\vskip 0.2in
\begin{center}
\centerline{\includegraphics[width=1.5\columnwidth]{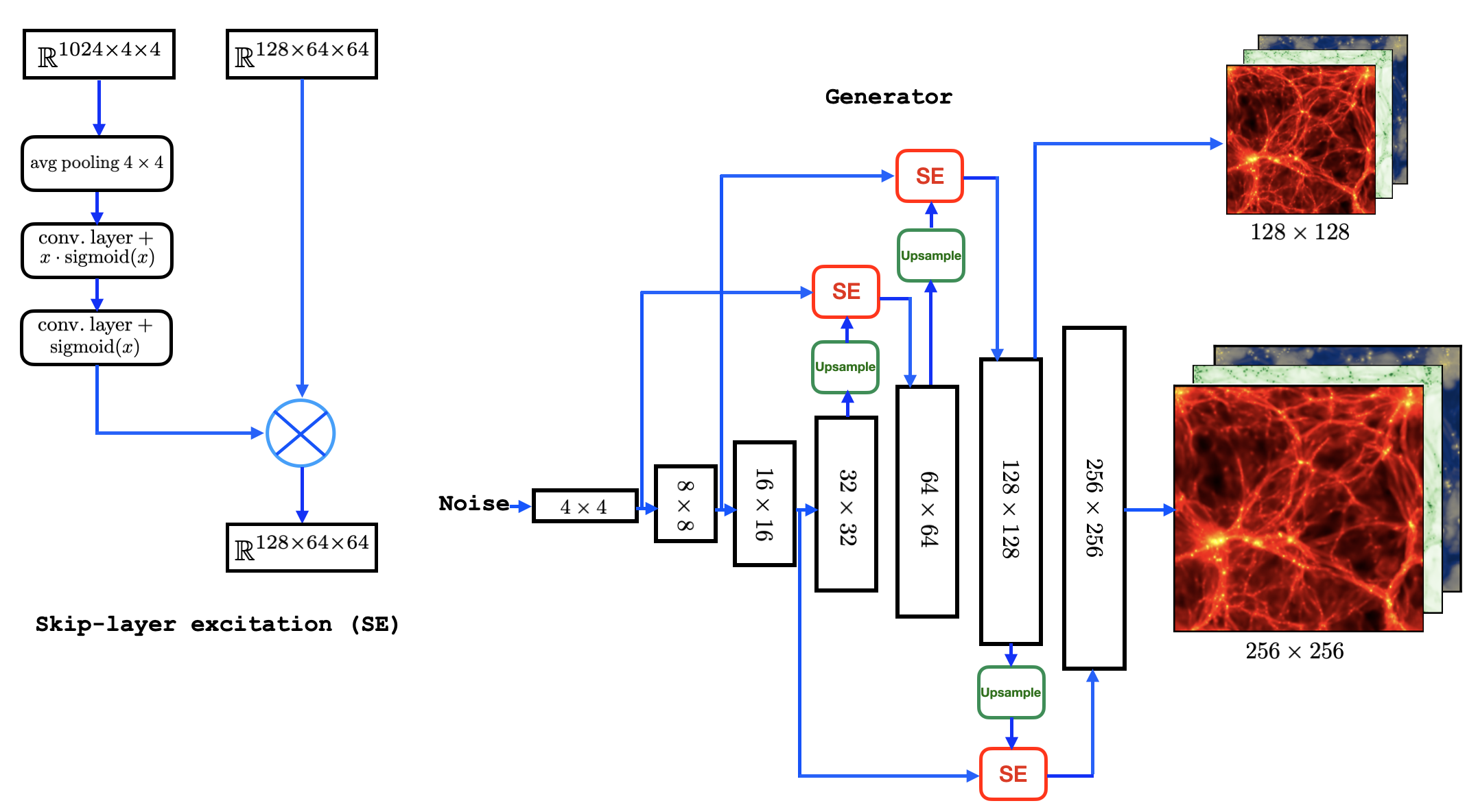}}
\caption{The schematic diagram of a Skip-layer excitation (SE) is shown on the \textit{left}, whereas that of the generator is presented on the \textit{right}. Two feature maps of different resolutions are fed into SE, and the one with a lower resolution goes through an average pooling and two convolutional layers before being channel-wise multiplied by the one with a higher resolution. The generator outputs a fake instance with the same dimensions as the data, i.e. $256\times 256 \times 3$, and the same fake instance but with lower dimensions, $128\times 128 \times 3$.} 
\label{fig:generator}
\end{center}
\end{figure*}
\begin{figure*}
\vskip 0.2in
\begin{center}
\centerline{\includegraphics[width=1.5\columnwidth]{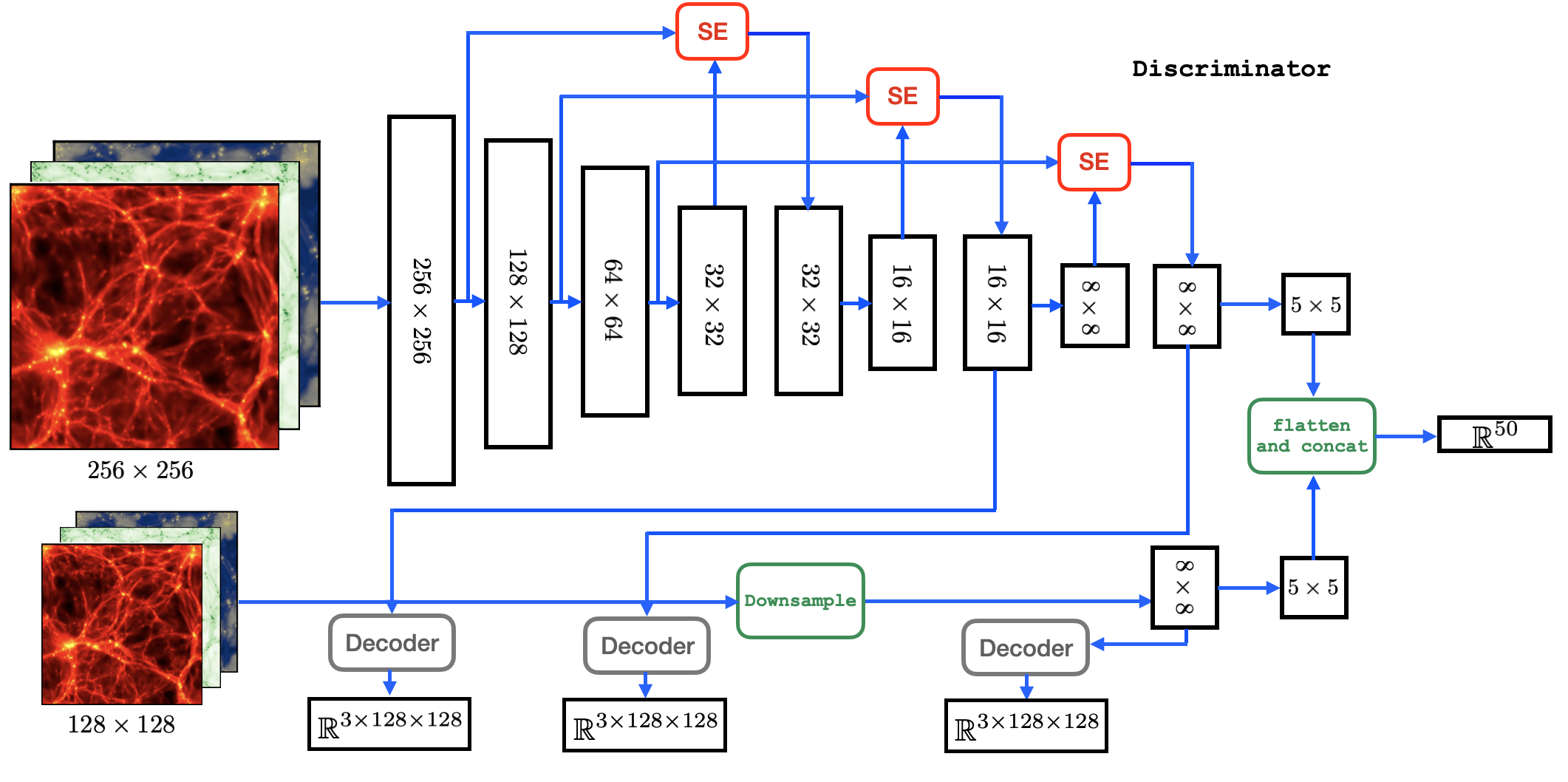}}
\caption{Schematic diagram of the discriminator which also uses SE. The discriminator takes both a lower and higher resolution instances as inputs. }
\label{fig:discriminator}
\end{center}
\vskip -0.2in
\end{figure*}
It is claimed in \cite{goodfellow2014generative} that the training is \textit{straightforward} when both models are multilayer perceptrons (MLP), i.e. mainly composed of dense layers. However, when local information is important and the number of features is large, such as in the case of an image, a convolution layer-based network is better suited for the task. To address this, for example, \cite{radford2015unsupervised} prescribed a GAN model in which both $D$ and $G$ use convolution layers and convolutional-transpose\footnote{Sometimes known as deconvolution layer} layers respectively.\\
We consider a variant of GANs described in \cite{liu2020towards} who introduced some novelties such that the generative model is capable of synthesizing high-resolution images while being trained with a relatively small number of examples. They prescribed a modified version of skip-layer connection \citep{he2016deep}, named \textit{Skip-Layer Excitation} (SLE) (see left panel in Figure~\ref{fig:generator}), which, like the original skip-layer, allows the gradient flow to be preserved through the layers. However, unlike the skip connection in \cite{he2016deep}, SLE does not require the two inputs to have the same dimension as the output results from channel-wise multiplications between the two inputs, according to \citep{liu2020towards}
\begin{equation}
    {\bm y} = \mathcal{F}({\bm x}_{\rm low}, {\bm W}) \cdot {\bm x}_{\rm high},  
\end{equation}
where ${\bm x}_{\rm low}$ and ${\bm x}_{\rm high}$ are two input feature maps with lower and higher resolutions respectively. $\mathcal{F}$ denotes a module with learnable weights ${\bm W}$ that operates on ${\bm x}_{\rm low}$. The generator \textit{G}  in our model contains three SLE modules that perform channel-wise multiplications of maps at $4\times4$, $8\times8$ and $16\times16$ resolutions with those of $64\times64$, $128\times128$,  and $256\times256$ resolutions respectively (see Figure~\ref{fig:generator}). It is noted that, and as shown in Figure~\ref{fig:generator}, \textit{G} outputs low  ($128\times 128 \times 3$) and high resolution ($256\times 256\times 3$) instances which are passed through the discriminator.  
Following the prescription in \cite{liu2020towards}, the discriminator \textit{D}, treated as an encoder, is trained with three decoders in order to enhance its ability to extract the salient features from the input maps (Figure~\ref{fig:discriminator}). The auto-encoder, which comprises \textit{D} and the decoders, is optimized using only real images by minimizing the reconstruction loss \citep{liu2020towards}
\begin{equation}\label{reconstruction}
    \mathcal{L}_{\rm recons} = \mathbb{E}_{{\bm f}\sim D({\bm x})}[||\mathcal{G}({\bm f}) - \mathcal{T}(x)||],
\end{equation}
where ${\bm f}$ denotes intermediate feature maps with resolutions $8\times8$ and $16\times16$ from \textit{D} (see Figure~\ref{fig:discriminator}), $\mathcal{G}$ is the decoding process and $\mathcal{T}$ represents a transformation on the real images $\bm x$, namely cropping and downsampling. As already mentioned, \textit{D} expects two inputs which are the same image with two different dimensions as shown in Figure~\ref{fig:discriminator}. 
In the case of generated instances, the discriminator processes the two outputs of dimensions $128\times 128 \times 3$ and $256\times 256\times 3$ from the generator to compute the generator loss. In the case of real data, both the lower resolution ($128\times 128 \times 3$), which is obtained from downsampling the data via interpolation, and the higher resolution ($256\times 256 \times 3$) are used to calculate both the discriminator and the reconstruction losses (Equation~\ref{reconstruction}).
For the adversarial training, we have that \citep{liu2020towards}
\begin{eqnarray}
&&\mathcal{L}_{D} = -\mathbb{E}_{{\bm x}\sim p_{\rm data}({\bm x})}[{\rm min}(0, -1 + D({\bm x}))]\nonumber\\
&& ~~~~ - \mathbb{E}_{{\bm z}\sim p_{\rm z}({\bm z})}[{\rm min}(0, -1 - D(G({\bm z})))] + \mathcal{L}_{\rm recons}
\end{eqnarray}
as the loss function for \textit{D} and 
\begin{equation}
    \mathcal{L}_{G} = -\mathbb{E}_{{\bm z}\sim p_{\rm z}({\bm z})}[D(G({\bm z}))],
\end{equation}
the loss for \textit{G}. It is worth noting that $\bm z$, which is a vector from latent space, is drawn from a normal distribution.
\begin{figure}[ht]
\vskip 0.2in
\begin{center}
\centerline{\includegraphics[width=\columnwidth]{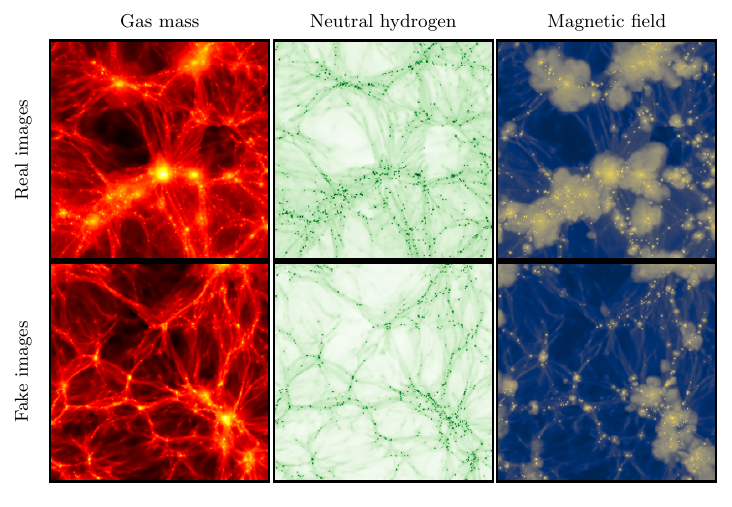}}
\caption{Comparison between CAMELS images (top row) and the generated images (bottom row) using unconditional GANs for randomly selected set of fields. The top row shows images of gas mass density (left), neutral hydrogen mass density (middle), and magnetic field strength (right) from a state-of-the hydrodynamic simulation from the CAMELS project. Note that the three images represent the same region in space. The bottom row shows each field of an instance output by our GAN model. From visual inspection, we can see that the model very well reproduces the morphological features of the different fields.}
\label{fig:real_fake_maps}
\end{center}
\vskip -0.2in
\end{figure}
\begin{figure}[ht]
\vskip 0.2in
\begin{center}
\centerline{\includegraphics[width=\columnwidth]{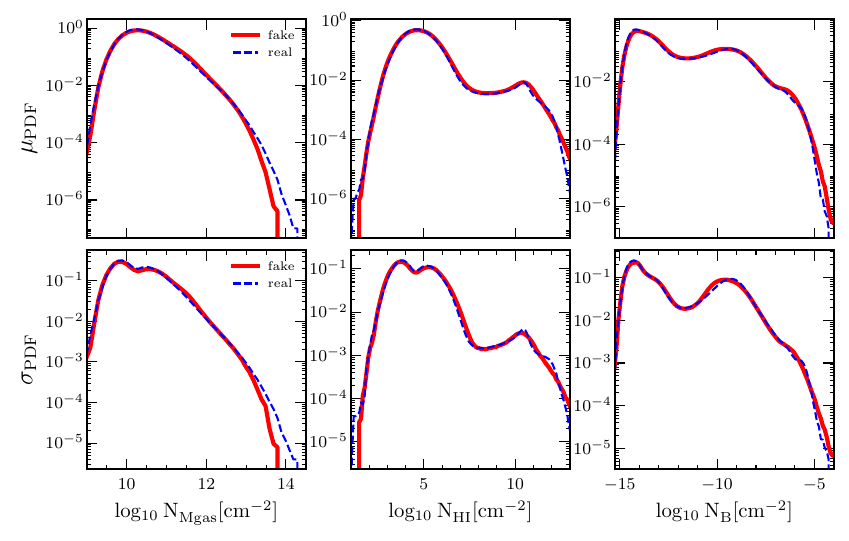}}
\caption{Comparison between CAMELS and unconditional GAN in terms of the average (top) and standard deviation (bottom) of the probability distribution function of pixel values (PDF) over 1,500 different realizations. We have computed the PDF for 1,500 unseen real images (red) and 1,500 generated images (blue) for each channel (or field). This plot shows the mean (top) and standard deviation (bottom) of the PDF for the three different fields. The agreement between the results of the simulated and generated images is good overall, however some statistical differences at the tails of the distributions can be noticed.}
\label{fig:pdf_real_fake}
\end{center}
\vskip -0.2in
\end{figure}
\begin{figure}
\vskip 0.2in
\begin{center}
\centerline{\includegraphics[width=\columnwidth]{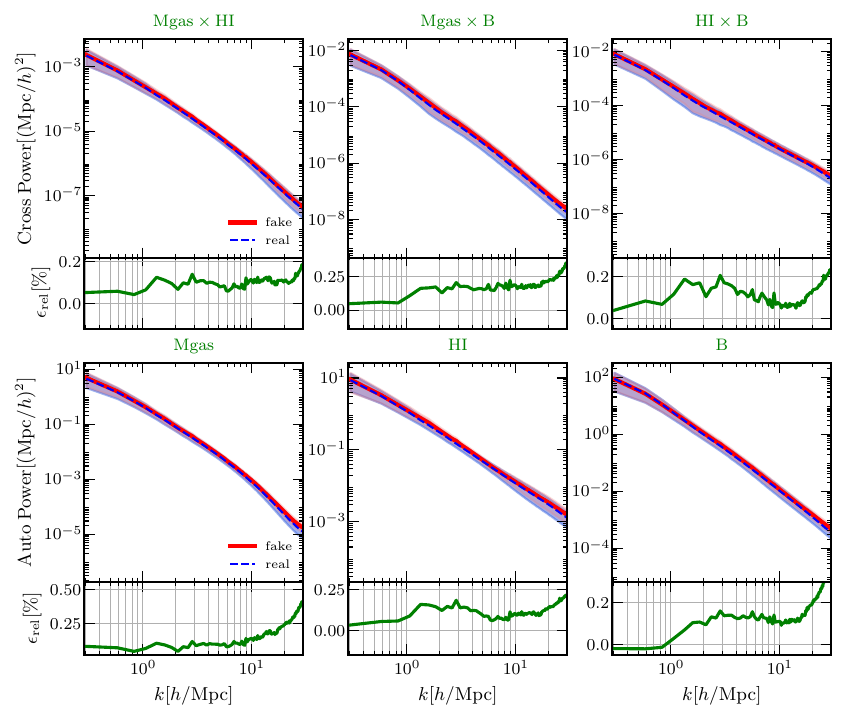}}
\caption{Similar to Figure~\ref{fig:pdf_real_fake} but for the cross-power spectrum (top row) and auto-power spectrum (bottom row). Relative error between the two power spectra (fake and real), denoted by solid green, is presented at the bottom part of each panel. As can be seen, the statistical agreement between the data and generated images up to the second moment is relatively good for this statistics. The title of each panel displays the considered field(s).}
\label{fig:power_spectra}
\end{center}
\vskip -0.2in
\end{figure}
\section{Data}\label{sec:data}
We make use of images from the CAMELS Multifield Dataset (CMD) \cite{CMD}. CMD contains hundreds of thousands of 2D maps generated from the output of state-of-the art hydrodynamic simulations of CAMELS \citep{villaescusa2021camels} for 13 different physical fields, e.g. dark matter, magnetic fields. Every image represents a region of dimensions $25\times25\times5~(h^{-1}{\rm Mpc})^3$ at $z=0$. In our analyses, we consider images comprising three different fields -- gas mass density (\verb|Mgas|), neutral hydrogen density (\verb|HI|), and magnetic fields magnitude (B) -- generated from the IllustrisTNG LH set \cite[see][for details]{villaescusa2021camels, pacodatarelease}. Each field contains 15,000 maps which were produced from varying the cosmology ($\Omega_{\rm m}$ and $\sigma_8$), the astrophysics ($A_{\rm SN1}$, $A_{\rm SN2}$, $A_{\rm AGN1}$, $A_{\rm AGN2}$), and the initial seed number. It is worth noting that the supernova feedback process is characterized by $A_{\rm SN1}$ and $A_{\rm SN2}$ which control the energy transfer and the wind speed respectively. $A_{\rm AGN1}$ and $A_{\rm AGN2}$, which encode the power injected and jet speed respectively, characterize the AGN feedback process. The multifields are produced by stacking 3 field images that show the same physical region: Mgas, HI, and B. Thus, each instance in our training data has $256\times256$ pixels$^2$ and contains three channels\footnote{Tensors of dimensions $256\times256\times 3$}. For training, we set the learning rate to 0.0002 and employ Adam optimizer. The model is trained with 10,000 multifield images for 700 epochs\footnote{Each epoch takes about 12 minutes.} in batches of 8 examples on a NVIDIA GeForce GTX 1080 Ti.
\begin{figure}[ht]
\vskip 0.2in
\begin{center}
\centerline{\includegraphics[width=\columnwidth]{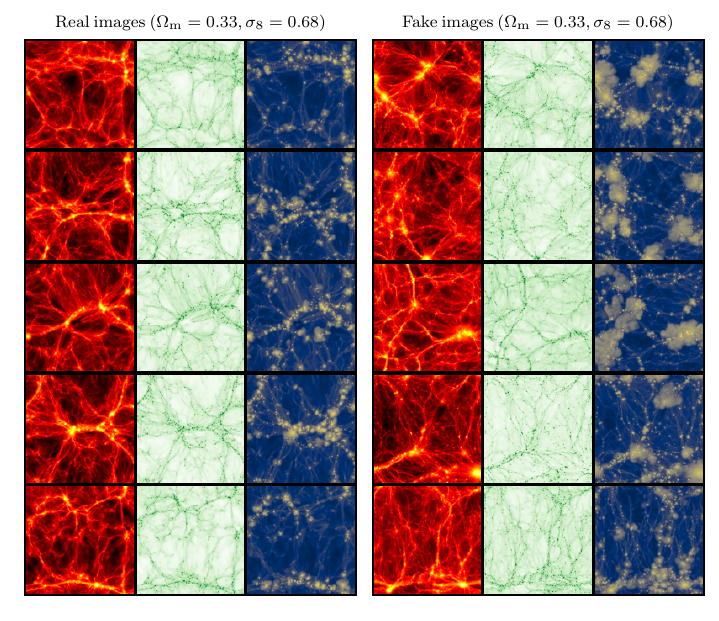}}
\caption{Left panel presents samples from the test data whose cosmology is known ($\Omega_{\rm m} = 0.33$, $\sigma_{8} = 0.68$). Conditioned by the same set of cosmological parameters, the images on the right panel have been generated by the conditional GAN. The color coding for the fields in each instance is the same as in Figure~\ref{fig:real_fake_maps}, i.e. red, green and blue refer to Mgas, HI and B respectively.}
\label{fig:real_fake_maps_cond}
\end{center}
\vskip -0.2in
\end{figure}

\begin{figure}[ht]
\vskip 0.2in
\begin{center}
\centerline{\includegraphics[width=\columnwidth]{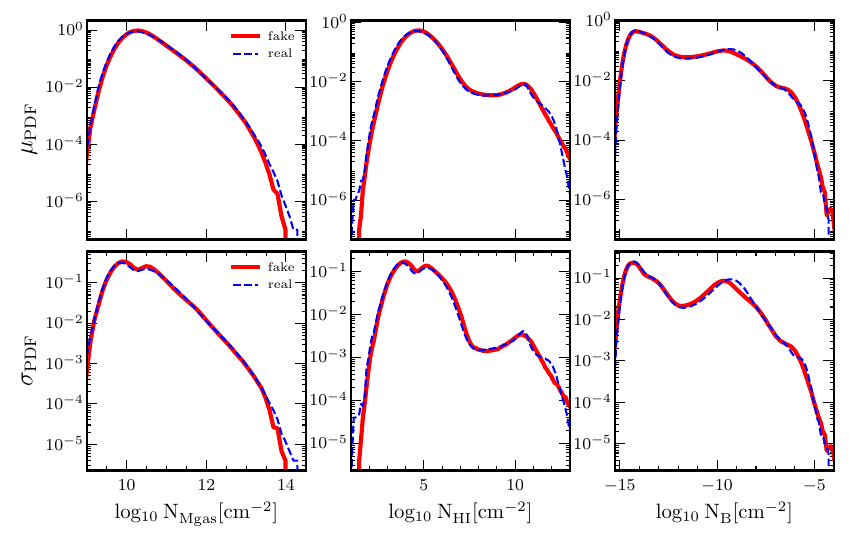}}
\caption{The mean and standard deviation of PDF of the generated maps conditioned by the cosmology ($\Omega_{\rm m}$ and $\sigma_{8}$) are shown at the top row and bottom row respectively. Each column correspond to a field.}
\label{fig:pdf_real_fake_cond}
\end{center}
\vskip -0.2in
\end{figure}

\begin{figure}
\vskip 0.2in
\begin{center}
\centerline{\includegraphics[width=\columnwidth]{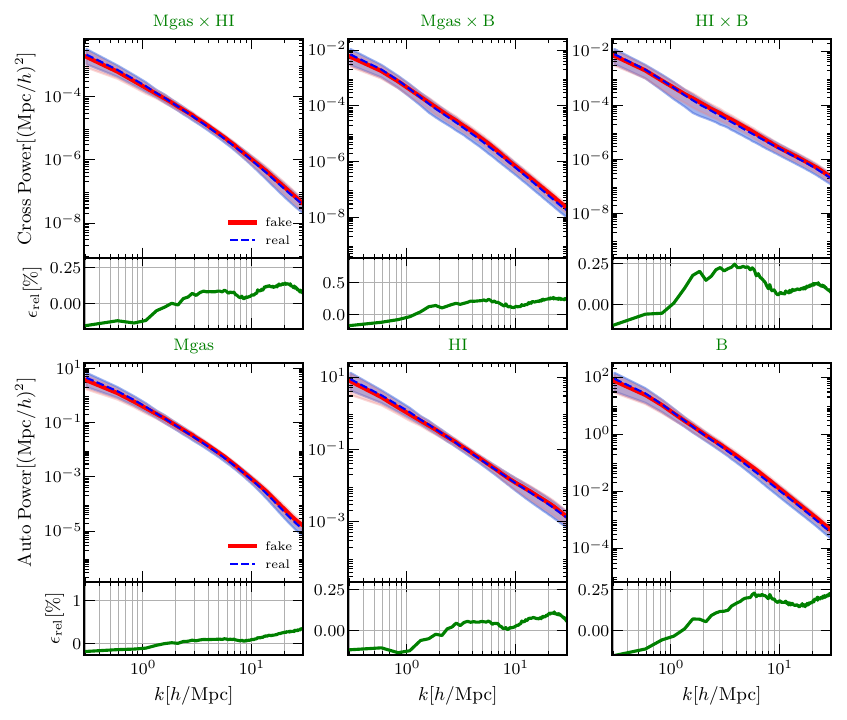}}
\caption{Similar to Figure~\ref{fig:power_spectra} but the results are obtained from conditioning the image generations on the two cosmological parameters ($\Omega_{\rm m}$ and $\sigma_{8}$).}
\label{fig:power_spectra_cond}
\end{center}
\vskip -0.2in
\end{figure}
\section{Results}\label{sec:results}
We now present the results obtained from the two scenarios in our analyses, i.e. unconditional and conditional GANs. The model performance is evaluated using probability distribution function (PDF) of the pixel values and power spectrum $P(k)$.

\subsection{Unconditional GAN}\label{subsec:un-gan}
In Figure~\ref{fig:real_fake_maps}, we compare an instance output by our model with a randomly drawn image from our test set which comprises 1,500 examples that are unseen by the model during training. The top row corresponds to real images\footnote{Which is also used to denote images from IllustrisTNG dataset.} and the bottom one shows the generated images\footnote{Which are images generated by the model.}. Each column in Figure~\ref{fig:real_fake_maps} is related to one field. Visually, the quality of the field maps in the generated instance is comparable with that of the real images from the test data. Morever, similar to the real instance, the morphological features -- voids, filaments, halos, and bubbles -- in each field map of the new example\footnote{In other words, the generated example.} correspond well with those of the other two channels (see bottom row in Figure~\ref{fig:real_fake_maps}). This is expected since all channels are supposed to represent the same physical region and indicates the ability of our model to produce images that mimic those of the training data. 

To assess the statistical properties of the maps generated by the model, we first use the PDF of the pixel intensities. By flattening each field map of an instance to an array, we compute the mean $\mu_{\rm PDF}$ and standard deviation $\sigma_{\rm PDF}$  of each channel of the 1,500 test data and generated instances. 
Figure~\ref{fig:pdf_real_fake} shows the comparison between the distributions of both real and new images. The top row shows the mean of the pixel value distribution, whereas the standard deviation is presented in the bottom row (see Figure~\ref{fig:pdf_real_fake}). Each column corresponds to results related to one field. We find that the generator is able to encode the salient features of each field, as evidenced by the overall agreement between the $\mu_{\rm PDF}$ of generated instances and that of CAMELS, despite the statistical differences at the tails of the distributions. The latter can be explained by the fact that high-valued pixels are rare in the maps, hence the induced bias. The consistency between the variability of the pixel distribution of both new and real examples (see $\sigma_{\rm PDF}$ at the bottom row of Figure~\ref{fig:pdf_real_fake}) implies that the model has learned a robust lower dimension representation of the data.

The metric we use to evaluate the clustering properties of the new images is the power spectrum $P(k)$ which is computed using  {\sc Pylians}\footnote{\url{https://pylians3.readthedocs.io}} \citep{villaescusa2018pylians}. There are two aspects of this evaluation: assessing if the clustering properties of the generated images of each field are consistent with their real counterparts by computing the average auto-power spectra ($\bar{P}^{\rm auto}(k)$) of both generated and real examples; and investigating the consistency between the three channels of the new instances and comparing it with that of the data, by computing the average cross-power spectra between channels ($\bar{P}^{\rm cross}(k)$). 
\begin{figure*}
\vskip 0.2in
\begin{center}
\centerline{\includegraphics[width=1.5\columnwidth]{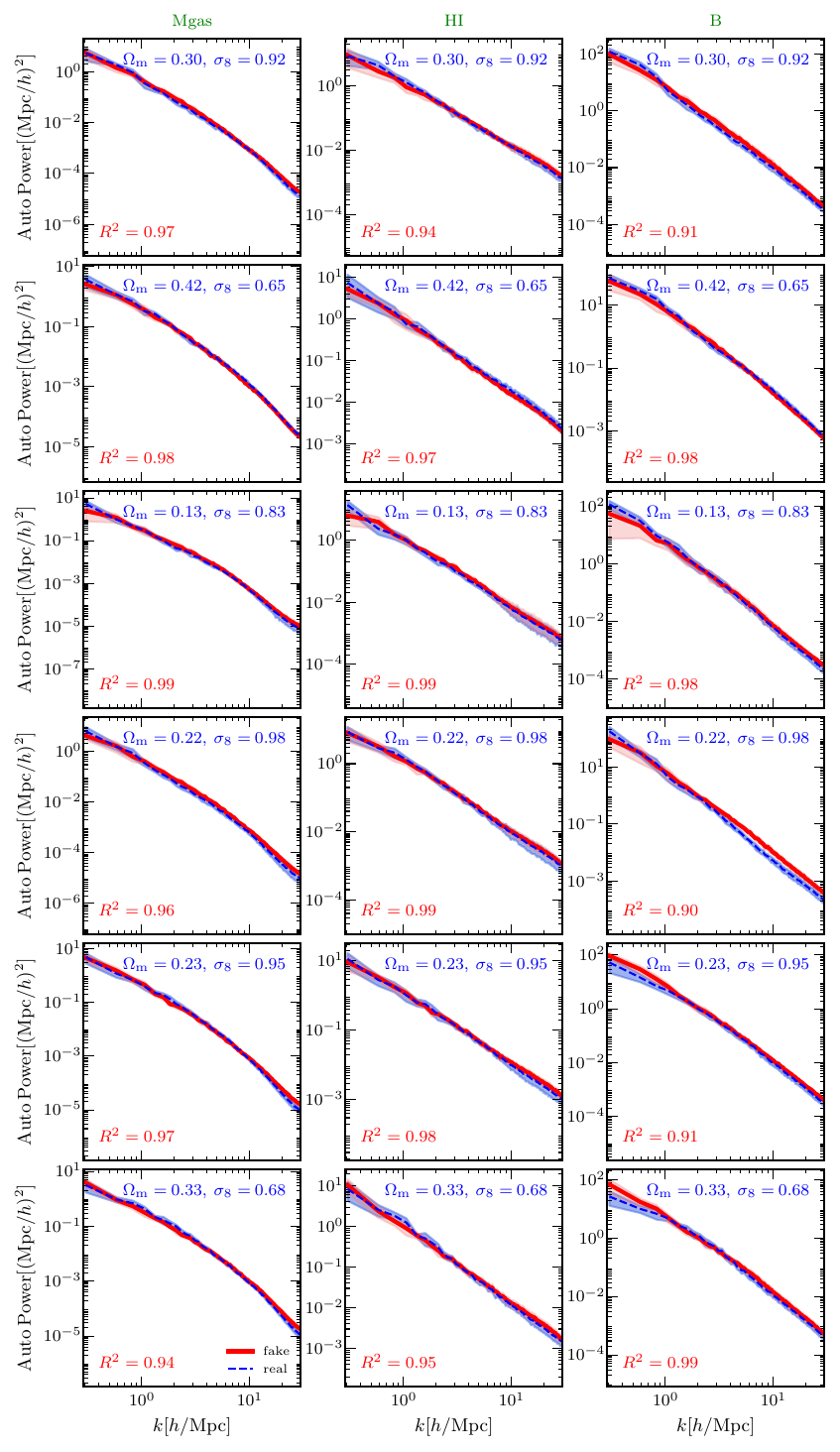}}
\caption{For each field, the average auto-power spectrum of the data (using 10 instances from the test set) are compared with that of the generated samples (using 10 GAN-generated samples). Each row corresponds to the power spectra of all fields for a given cosmology ($\Omega_{\rm m}$ and $\sigma_{8}$). The coefficient of determination $R^{2}$ between the  predicted and real power spectra of each field is computed and shown at the bottom left of each panel.}
\label{fig:all_autopk_cond}
\end{center}
\vskip -0.2in
\end{figure*}
\begin{figure*}
\vskip 0.2in
\begin{center}
\centerline{\includegraphics[width=1.5\columnwidth]{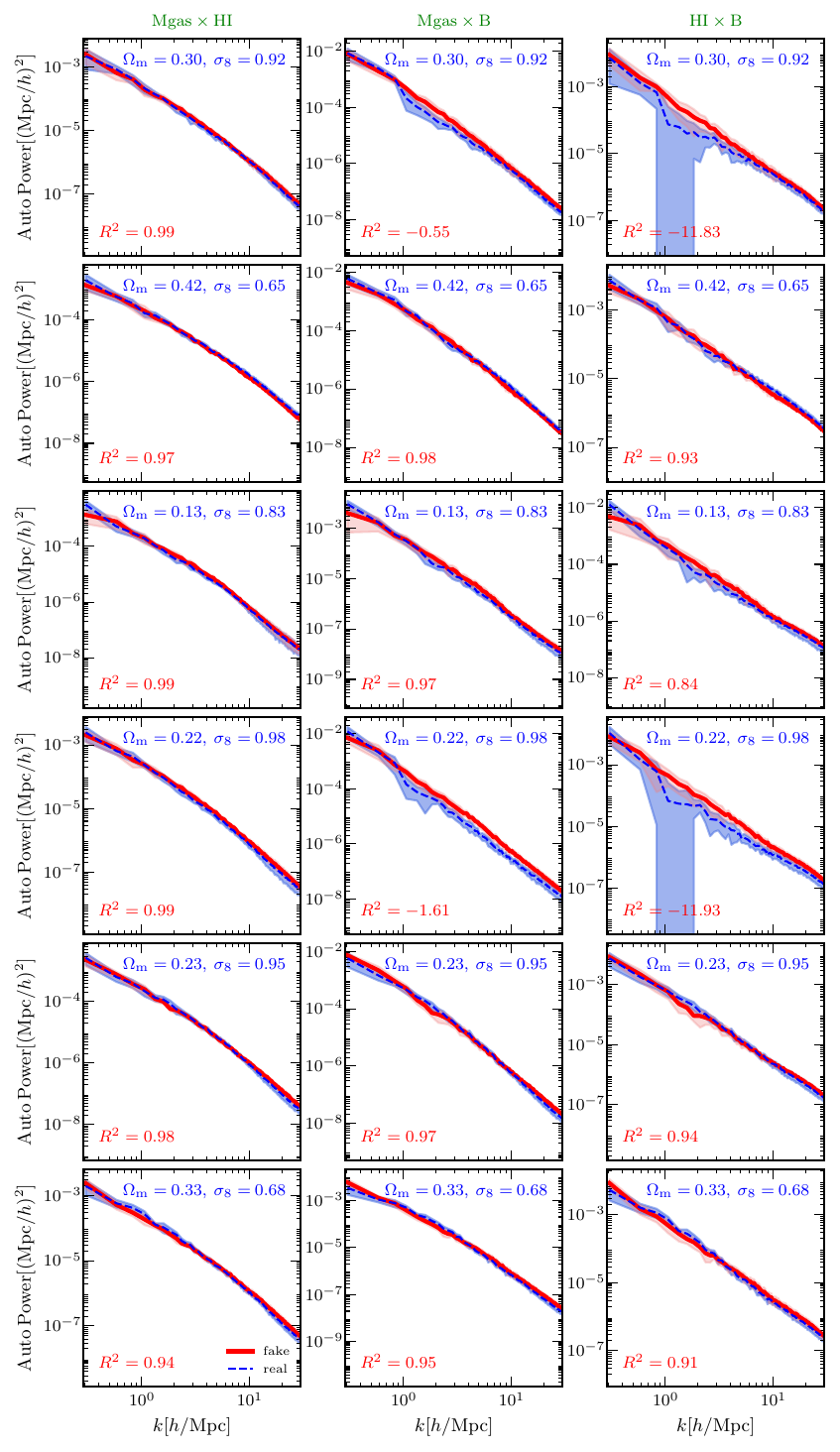}}
\caption{Similar to what is shown in Figure~\ref{fig:all_autopk_cond} but for the cross-correlation between fields of both real and fake images. The average cross-power spectra of both the data and new images are computed using 10 instances from $\mathbb{S}^{\rm cosmo}$ subset, 10 GAN-generated samples respectively.}
\label{fig:all_crosspk_cond}
\end{center}
\vskip -0.2in
\end{figure*}
We present the average auto and cross-power spectra of the images at the bottom and top rows of Figure~\ref{fig:power_spectra} respectively. In each panel, the solid red and dashed blue lines indicate the average power spectra of the generated and simulated images respectively; and the red and blue shaded areas denote the standard deviations of the power spectra of the new and real images respectively. The solid green line at the bottom of each panel represents the relative difference between the fake and real average power spectra at each $k$-mode $\left(\frac{\bar{P}_{\rm fake}(k)}{\bar{P}_{\rm real}(k)} - 1\right)$. The good agreement between the mean auto-power spectra, and also between the corresponding standard deviations, in each field is indicative of the fact that the generator is capable of producing images whose clustering properties are consistent with those of the data. However, it appears that modeling the fluctuations of each field is more challenging on small scales, as denoted by the larger discrepancies on scales $k > 10\>h/{\rm Mpc}$ (see also the solid green line at the bottom of each panel at the bottom row of Figure~\ref{fig:power_spectra}). 
The results in Figure~\ref{fig:power_spectra} top panels suggest that the average cross-correlation signals between channels in the generated instances agree with those of the data, implying that all channels in each example generated by the model represent the same physical region as expected. Similar to the case of auto-power spectra, the relative difference between the predicted and real average cross-power spectra is more important with increasing wavenumber $k$. The agreement between the blue and red shaded areas (top panels in Figure~\ref{fig:power_spectra}) demonstrates the ability of the generator to capture the variability of the signals at each $k$-mode, denoting that the model generalizes well. Technically, our results can't be directly compared with those obtained in \cite{tamosiunas2021investigating} as their scale of interest is $k \le 1\>h/{\rm Mpc}$. However the relative error on the auto-power spectrum of each field on scales $k \le 1\>h/{\rm Mpc}$ (see bottom row of Figure~\ref{fig:power_spectra}) is comparable to that obtained in \cite{tamosiunas2021investigating} (see the top left panel in their Figure~15).

\subsection{Conditional GAN}\label{subsec:con-gan}
In light of the results presented in \S~\ref{subsec:un-gan}, it can be said that the model is capable of producing new instances which exhibit statistical properties closely similar to those of the data. While very promising, the underlying cosmology and astrophysics are unknown in the unconditional case. When exploring parameter space which requires accelerating cosmological simulations, we wish to obtain the resulting map that corresponds to the cosmological and astrophysical parameters at the input. However, the astrophysical parameters are highly degenerate and our aim, as stated in \S~\ref{sec:Intro}, is to constrain cosmology while marginalizing over the poorly constrained astrophysics.
Therefore, in our analyses, we condition the GAN model on the two cosmological parameters ($\Omega_{\rm m}$ and $\sigma_{8}$) that the maps are most sensitive to in general. To do so, we retrain the model using the same set of hyperparameters as that in the unconditional case (see \S~\ref{subsec:un-gan}), but both the generator and discriminator are dependent on $\Omega_{\rm m}$ and $\sigma_{8}$. Conditioning the generator consists of simply concatenating the cosmological parameters (array of length 2) to the prior vector at its input. Implementing the condition in the discriminator amounts to passing the cosmological parameters through a dense layer of 4096 units whose output is reshaped to a tensor $64\times64$. The latter dimension is first upsampled to match the channel dimension ($256\times256$) via interpolation and then concatenated to the input, yielding a new input comprising four channels.
In Figure~\ref{fig:real_fake_maps_cond} on the left panel, we show  samples from the test data that have the same underlying cosmology\footnote{We note that those samples from the test set have the same set of astrophysical parameters as well.} which has also been used to condition the generation of the new images on the right panel. The diversity of the GAN-generated images (see Figure~\ref{fig:real_fake_maps_cond}), although corresponding to the same cosmology, is noticeable. This points toward the fact the GAN model does not suffer from mode collapse. 
Consistent with the results obtained in \S~\ref{subsec:un-gan}, the quality of the generated images is on a par with that of simulated ones, and visually all channels of each new instance appear to show the same physical region. Following the performance tests carried out in the unconditional case, the PDF and power spectra are computed using 1,500 unseen samples from the test set and 1,500 generated samples. Overall, the estimated mean $\mu_{\rm PDF}$ and standard deviation $\sigma_{\rm PDF}$ of pixel value distribution indicate that the generated images conditioned on the cosmology exhibit topology which agrees well with that of the data, as shown in Figure~\ref{fig:pdf_real_fake_cond}. Results show that the conditional GAN has also learned to capture the clustering properties of each field from the training data, as indicated by the overall agreement between the auto-power spectra of generated maps and those of real maps for each field (see Figure~\ref{fig:power_spectra_cond} bottom row). The fact that cross-correlation signals between channels of the generated instances are consistent with their real counterparts (see Figure~\ref{fig:power_spectra_cond} top row) implies that all the channels in each GAN-generated example are consistent with one another and depict the same physical region. 

In general, based on the metrics considered in our analyses, both unconditional and conditional GAN perform equally well, but we further investigate how well the new images are constrained by the cosmology ($\Omega_{\rm m}$, $\sigma_{8}$). Six subsets of ten images are selected from the holdout set. The images in each subset correspond to the same cosmology, characterized by the pair $\{\Omega_{\rm m},\sigma_{8}\}$, which is also used to condition the generations of six subsets of ten new images. In other words, in the entire subset which we name $\mathbb{S}^{\rm cosmo}$\footnote{$\mathbb{S}^{\rm cosmo}$ contains $6\times$ 10 CAMELS data, and $6\times$ 10 generated data, giving 1200 samples in total.} for clarity, each pair $\{\Omega_{\rm m},\sigma_{8}\}$ has 20 corresponding images in which half is drawn from the test set and the other half is produced by the model. To assess the performance of the model in this test, like the one conducted in \cite{hassan2021hiflow}, the average ($\bar{P}(k)$) and standard deviation ($\sigma_{P}(k)$)  power spectra  are computed for each set of cosmological parameters using the ten samples of each subset (real/new). In Figure~\ref{fig:all_autopk_cond} we compare the predicted mean auto-power spectra $\bar{P}_{\rm fake}(k)$ (solid red line) with the real ones $\bar{P}_{\rm real}(k)$  (dashed blue line) for different sets of cosmological parameters. The field in each panel of each row corresponds to the same cosmology whose parameters are shown in blue on the top right corner. The standard deviation of the power spectra at each $k$-bin is denoted by red and blue shaded areas for the GAN-generated and CAMELS instances respectively. Coefficient of determination $R^{2}$ is used to check the consistency between the power spectra and is given by
\begin{equation}\label{r2score}
    R^{2} = 1 - \frac{\sum_{i = 1}^{n}\left(\bm{y}_{i}-\hat{\bm{y}}_{i}\right)^{2}}{\sum_{i = 1}^{n}\left(\bm{y}_{i}-\bar{\bm{y}}_{i}\right)^{2}},
\end{equation}
where $n$, $\bm{y}_{i}$ (or $\bar{P}_{\rm real, i}$), $\hat{\bm{y}}_{i}$ (or $\bar{P}_{\rm fake, i}$) are the number of $k$-bins, data and generated images power spectra at each $k$-bin respectively, and we have that $\bar{\bm{y}}_{i}=(1/n)\sum_{i = 1}^{n}\bm{y}_{i}$. 
For more accurate estimate of $R^{2}$, we only select $k$-bins with more than ten modes. In other words, our calculations are restricted to $k \ge 1 h/{\rm Mpc}$.
In general, for any of the cosmologies and in each field, the clustering properties of CAMELS and those of the generated data are consistent with each other, as evidenced by all $R^{2}$ values $\ge 0.9$ (see Figure~\ref{fig:all_autopk_cond}). This suggests that the image generations can be constrained reasonably well by the cosmological parameters. We also inspect the consistency between the field maps in each GAN-produced sample when the cosmology is fixed. In Figure~\ref{fig:all_crosspk_cond} we present the average cross-correlation between channels of all instances (CAMELS/GAN-generated) in $\mathbb{S}^{\rm cosmo}$. Like in Figure~\ref{fig:all_autopk_cond}, the results in all panels of each row are related to a given cosmology. Each column corresponds to the same cross-correlation between two fields. Overall, for each set of cosmological parameters in this test, the GAN model is able to predict reasonably well the cross-correlations between the different fields (Mgas$\times$HI, Mgas$\times$B and HI$\times$B) of CAMELS, with most of $R^{2}$ values $\ge 0.84$. However there are cases where discrepancy between two power spectra (of CAMELS and GAN-generated images) is relatively important, as indicated by negative values of $R^{2}$. Predicting the cross power spectra of both Mgas$\times$B and HI$\times$B for $\{\Omega_{\rm m}=0.30,\sigma_{8} = 0.92\}$ and $\{\Omega_{\rm m}=0.22,\sigma_{8} = 0.98\}$ cosmologies seems challenging to the generative model. This could be potentially explained by the unconstrained stellar feedbacks $A_{\rm SN1}$ and $A_{\rm SN2}$ which have a non-negligible effect on the PDFs of HI and B. 
\begin{table}
 \centering
 \begin{tabular}{|c|c|c|c|}
  \hline
   &  HI & Mgas-HI & Mgas-HI-B \\
  \hline
  $\Omega_{\rm m}$  & 0.80 (0.99) & 0.92 (0.99)& 0.96 (0.99)\\[2pt]
  $\sigma_{8}$  & -0.11 (0.92) & 0.63 (0.94)& 0.83 (0.96)\\[2pt]
  \hline
 \end{tabular}
\caption{$R^{2}$ obtained from predicting the cosmological parameters on the images generated by the conditional generator in three different setups, which depend on the number of channels at the input of a neural network trained in \citep{andrianomena2022predictive}. The values in round brackets denote the performance of the CNNs in \cite{andrianomena2022predictive}.}
 \label{tab:params-pred}
\end{table}
\subsection{Extracting cosmological parameters from generated maps}\label{subsec:extract-params}
We further analyze the features learnt by the GANs by extracting the underlying cosmology of the generated maps.
To this end, following \cite{andrianomena2022predictive}, we consider three different setups: predictions on neutral hydrogen maps, predictions on instances that comprise two channels (gas density and neutral hydrogen), and predictions on instances with three channels (gas density, neutral hydrogen, and magnetic fields). 
In each scenario, we make use of the trained network models in \cite{andrianomena2022predictive}, which are convolutional neural networks (CNN). To assess the performance of the regressors, we again consider the coefficient of determination ($R^{2}$) which indicates both the ability of the regression model to account for the variability in the predicted parameters and the strength of the correlation between the ground truth and prediction. It is noted that we choose the labels ($\Omega_{\rm m}$, $\sigma_{8}$) of the test set, which is used to assess the performance of the trained CNNs in \cite{andrianomena2022predictive}, to condition the generation of 1,500 new instances. Depending on the setup (HI, Mgas-HI or Mgas-HI-B), only the channel of interest is selected from the new examples, e.g. for HI setup, only the HI maps from the generated test sample is considered. The performance of the CNN model on the fake images can be directly compared with its performance on the CAMELS data reported in \cite{andrianomena2022predictive}. Each column in Table~\ref{tab:params-pred} corresponds to the resulting $R^{2}$ for both $\Omega_{\rm m}$ and $\sigma_{8}$ in each scenario. 
On the one hand, it is clear that, irrespective of the case, the CNN models are able to recover the matter density with relatively good accuracy, as indicated by $R^{2} \ge 0.80$ in all setups. 
In the case of three channel inputs (Mgas-HI-B), the ability of the CNN model to constrain $\Omega_{\rm m}$ from the GAN-generated images is comparable to its performance on the CAMELS data, as evidenced by $R^{2}$ = 0.96, 0.99 (see Table~\ref{tab:params-pred}) corresponding to the matter density constraint achieved on the new images and CAMELS respectively. The $\Omega_{\rm m}$ constraint obtained from the generated images is still reasonably good on the two channel (Mgas-HI) setup ($R^{2} = 0.92$), but is relatively worse on the generated single channel HI maps ($R = 0.80$). This points to both how well the deep regression model in \cite{andrianomena2022predictive} generalizes, i.e. its predictive power is such that it is able to identify salient features of the fake maps in order to infer the matter density, and the fact that the generative model has learnt good representations of the CAMELS data to be able to mimic it. 
On the other hand, retrieving $\sigma_{8}$ from the new images seems to be more challenging. The CNN model fails to extract $\sigma_{8}$ from the fake HI maps ($R^{2} = -0.11$) but its performance on the CAMELS-like data improves with an increasing number of channels at the input, reaching $R^{2} = 0.83$ on the Mgas-HI-B setup. This trend (Table~\ref{tab:params-pred}), in which the constraint on $\sigma_{8}$ gets tighter with more information at the input, was also noticed in \cite{andrianomena2022predictive}.

\section{\label{conclusions}Conclusions}
In this work we have shown the ability of a generative model to produce multifield images whose statistical properties are consistent with those of the training data. We resort to generative adversarial networks (GANs), which consist of two networks, a generator and a discriminator that are trained adversarially. We have made use of CAMELS multifield dataset \citep{CMD} and considered three different fields obtained from CAMELS IllustrisTNG simulations: gas density (Mgas), neutral hydrogen density (HI), and magnetic field magnitudes (B). Each instance in our training set, which comprises 10,000 examples in total, consists of three channels (Mgas-HI-B), each with a resolution of $256\times256$ pixel$^2$. We have explored one scenario where the model is not conditioned on any parameters and another one in which the generated instances are constrained by the cosmological parameters $\Omega_{\rm m}$ and $\sigma_{8}$. To assess the quality of the new examples generated by the GAN in both scenarios, we employ probability distribution function (PDF) of the pixel values and auto/cross-power spectrum. We have further tested the quality of the generated images whose cosmology is known by employing trained CNN models to recover their corresponding cosmological parameters. We summarize our findings as follows:
\begin{itemize}
    \item Irrespective of the case (unconditional or conditional), the quality of the images generated by the GAN considered in our study is comparable with that of the training data, exhibiting all the morphological features, e.g. filaments (Figures~\ref{fig:real_fake_maps} and \ref{fig:real_fake_maps_cond}). 
    \item We have found that the average of PDF $\mu_{\rm PDF}$ and standard deviation $\sigma_{\rm PDF}$ of the generated maps for each field are in good agreement with the those of the data, demonstrating that the generator has encoded a robust representation of the key features of the training data. Both the unconditional (Figure~\ref{fig:pdf_real_fake}) and conditional (Figure~\ref{fig:pdf_real_fake_cond}) models perform equally well.
    \item The generative model has learned the clustering properties of the training data. For each field, we have found that the average auto/cross-power spectrum of the new maps is consistent with that of the real maps in general, with a discrepancy of $\le 25\%$ up to $k \sim 10\>h/{\rm Mpc}$ in both unconditional and conditional cases. The second moments of the two power spectra also agree well.
    In all scenarios, the cross-correlation signal between channels in the GAN-generated examples is consistent with the training data on average. This indicates that all three maps in each new instance represent the same physical region (Figures~\ref{fig:power_spectra} and \ref{fig:power_spectra_cond}). 
    \item Provided the cosmology, we compare the predicted power spectra (auto/cross) with the real ones related to test data for each/cross field. It is found that for each cosmology considered in the test, the auto-correlations of the CAMELS and those of the GAN-generated data are in good agreement overall. However, the average cross-correlation signals 
    of some new images deviate from that of the data. The unconstrained supernova feedback ($A_{\rm SN1}$, $A_{\rm SN2}$) can potentially account for that.
    \item For the cosmological parameter recovery test on the GAN-generated data, the matter density is retrieved to a good accuracy in both Mgas-HI and Mgas-HI-B setups, with $R^{2}\ge 0.92$, whereas inferring $\sigma_{8}$ is a bit more complicated in that the combined information from the three fields, i.e. Mgas-HI-B setup, is required to get the best constraint on that parameter, corresponding to $R^{2} = 0.83$. It can be inferred from this test on the generated examples that the GAN model does not suffer from mode collapse which, amongst other factors, makes training adversarial networks challenging. As a whole, we argue that the capability of our generative model is very promising. 
\end{itemize}
We have only considered Mgas, HI and B in our analyses but the GAN model in this study can also be trained to reproduce other fields in the CAMELS project. The tool presented in the work can be used to help speed up map generation in IllustrisTNG hydrodynamic simulation, for instance within both the context of exploring parameter space which is computationally expensive and simulation based inference.  At this stage, image generation is only conditioned on cosmology. An improvement for future work will be to further condition the generative model on astrophysics as soon as a more refined and accurate models of subgrid astrophysics become available. 
The developed multifield  emulator in this work produces pure theoretically predicted quantities. To convert our predictions into mock observations, one needs to assume a particular experiment and survey design. This involves adding thermal noise, foreground cleaning and adjusting the angular resolution, to name a few instrumental effects (e.g. see \cite{hassan2019identifying}). However, most of the intensity mapping surveys are currently under development where the experiment/survey design is constantly being refined. Hence, we leave to interested user and future works to perform these additional postprocessing, once the setup of these large-scale instruments is finalized. In addition, our emulator is fast, and can easily act as an efficient multi-field simulator in Simulation Based Inference (SBI) pipelines, which we defer to future works when real observations become available, although the out-of-distribution challenge persists (e.g. see \cite{gondhalekar2023towards}). The developed emulator can be modified to generate 3D lightcones, which would enable accurate cross-correlation analysis with other experiments (such as LSS), which we leave for future works once the required computational resources including memory requirement are accessible. Another important aspect of our conditional model, which simultaneously generates three field maps corresponding to a cosmology, is its ability to infer both the gas density and magnetic field amplitudes, once the cosmological parameters have been extracted from neutral hydrogen maps.

It is worth noting that the magnetic field is correlated with the pressure, which is strongly correlated with the Star Formation Rate (SFR, see \cite{ostriker2022pressure}). It might be possible to infer the large-scale distribution of star formation from magnetic fields following some modelling assumptions, which we leave to future works.

\begin{acknowledgments}
SA acknowledges financial support from the {\it South African Radio Astronomy Observatory} (SARAO). SH acknowledges support for Program number HST-HF2-51507 provided by NASA through a grant from the Space Telescope Science Institute, which is operated by the Association of Universities for Research in Astronomy, incorporated, under NASA contract NAS5-26555. The CAMELS project is supported by the Simons Foundation and NSF grant AST 2108078.  
\end{acknowledgments}





\bibliography{refs}

\begin{thebibliography}{41}%
\makeatletter
\providecommand \@ifxundefined [1]{%
 \@ifx{#1\undefined}
}%
\providecommand \@ifnum [1]{%
 \ifnum #1\expandafter \@firstoftwo
 \else \expandafter \@secondoftwo
 \fi
}%
\providecommand \@ifx [1]{%
 \ifx #1\expandafter \@firstoftwo
 \else \expandafter \@secondoftwo
 \fi
}%
\providecommand \natexlab [1]{#1}%
\providecommand \enquote  [1]{``#1''}%
\providecommand \bibnamefont  [1]{#1}%
\providecommand \bibfnamefont [1]{#1}%
\providecommand \citenamefont [1]{#1}%
\providecommand \href@noop [0]{\@secondoftwo}%
\providecommand \href [0]{\begingroup \@sanitize@url \@href}%
\providecommand \@href[1]{\@@startlink{#1}\@@href}%
\providecommand \@@href[1]{\endgroup#1\@@endlink}%
\providecommand \@sanitize@url [0]{\catcode `\\12\catcode `\$12\catcode
  `\&12\catcode `\#12\catcode `\^12\catcode `\_12\catcode `\%12\relax}%
\providecommand \@@startlink[1]{}%
\providecommand \@@endlink[0]{}%
\providecommand \url  [0]{\begingroup\@sanitize@url \@url }%
\providecommand \@url [1]{\endgroup\@href {#1}{\urlprefix }}%
\providecommand \urlprefix  [0]{URL }%
\providecommand \Eprint [0]{\href }%
\providecommand \doibase [0]{https://doi.org/}%
\providecommand \selectlanguage [0]{\@gobble}%
\providecommand \bibinfo  [0]{\@secondoftwo}%
\providecommand \bibfield  [0]{\@secondoftwo}%
\providecommand \translation [1]{[#1]}%
\providecommand \BibitemOpen [0]{}%
\providecommand \bibitemStop [0]{}%
\providecommand \bibitemNoStop [0]{.\EOS\space}%
\providecommand \EOS [0]{\spacefactor3000\relax}%
\providecommand \BibitemShut  [1]{\csname bibitem#1\endcsname}%
\let\auto@bib@innerbib\@empty
\bibitem [{\citenamefont {National Academies~of Sciences}\ \emph
  {et~al.}(2021)\citenamefont {National Academies~of Sciences} \emph
  {et~al.}}]{NAP26141}%
  \BibitemOpen
  \bibfield  {author} {\bibinfo {author} {\bibfnamefont {M.}~\bibnamefont
  {National Academies~of Sciences}, \bibfnamefont {Engineering}} \emph
  {et~al.},\ }\href@noop {} {\emph {\bibinfo {title} {Pathways to Discovery in
  Astronomy and Astrophysics for the 2020s}}}\ (\bibinfo {year}
  {2021})\BibitemShut {NoStop}%
\bibitem [{\citenamefont {Predehl}\ \emph {et~al.}(2021)\citenamefont
  {Predehl}, \citenamefont {Andritschke}, \citenamefont {Arefiev},
  \citenamefont {Babyshkin}, \citenamefont {Batanov}, \citenamefont {Becker},
  \citenamefont {B{\"o}hringer}, \citenamefont {Bogomolov}, \citenamefont
  {Boller}, \citenamefont {Borm} \emph {et~al.}}]{predehl2021erosita}%
  \BibitemOpen
  \bibfield  {author} {\bibinfo {author} {\bibfnamefont {P.}~\bibnamefont
  {Predehl}}, \bibinfo {author} {\bibfnamefont {R.}~\bibnamefont
  {Andritschke}}, \bibinfo {author} {\bibfnamefont {V.}~\bibnamefont
  {Arefiev}}, \bibinfo {author} {\bibfnamefont {V.}~\bibnamefont {Babyshkin}},
  \bibinfo {author} {\bibfnamefont {O.}~\bibnamefont {Batanov}}, \bibinfo
  {author} {\bibfnamefont {W.}~\bibnamefont {Becker}}, \bibinfo {author}
  {\bibfnamefont {H.}~\bibnamefont {B{\"o}hringer}}, \bibinfo {author}
  {\bibfnamefont {A.}~\bibnamefont {Bogomolov}}, \bibinfo {author}
  {\bibfnamefont {T.}~\bibnamefont {Boller}}, \bibinfo {author} {\bibfnamefont
  {K.}~\bibnamefont {Borm}}, \emph {et~al.},\ }\bibfield  {title} {\bibinfo
  {title} {The erosita x-ray telescope on srg},\ }\href@noop {} {\bibfield
  {journal} {\bibinfo  {journal} {Astronomy \& Astrophysics}\ }\textbf
  {\bibinfo {volume} {647}},\ \bibinfo {pages} {A1} (\bibinfo {year}
  {2021})}\BibitemShut {NoStop}%
\bibitem [{\citenamefont {Dewdney}\ \emph {et~al.}(2009)\citenamefont
  {Dewdney}, \citenamefont {Hall}, \citenamefont {Schilizzi},\ and\
  \citenamefont {Lazio}}]{dewdney2009square}%
  \BibitemOpen
  \bibfield  {author} {\bibinfo {author} {\bibfnamefont {P.~E.}\ \bibnamefont
  {Dewdney}}, \bibinfo {author} {\bibfnamefont {P.~J.}\ \bibnamefont {Hall}},
  \bibinfo {author} {\bibfnamefont {R.~T.}\ \bibnamefont {Schilizzi}},\ and\
  \bibinfo {author} {\bibfnamefont {T.~J.~L.}\ \bibnamefont {Lazio}},\
  }\bibfield  {title} {\bibinfo {title} {The square kilometre array},\
  }\href@noop {} {\bibfield  {journal} {\bibinfo  {journal} {Proceedings of the
  IEEE}\ }\textbf {\bibinfo {volume} {97}},\ \bibinfo {pages} {1482} (\bibinfo
  {year} {2009})}\BibitemShut {NoStop}%
\bibitem [{\citenamefont {Weltman}\ \emph {et~al.}(2020)\citenamefont
  {Weltman}, \citenamefont {Bull}, \citenamefont {Camera}, \citenamefont
  {Kelley}, \citenamefont {Padmanabhan}, \citenamefont {Pritchard},
  \citenamefont {Raccanelli}, \citenamefont {Riemer-S{\o}rensen}, \citenamefont
  {Shao}, \citenamefont {Andrianomena} \emph
  {et~al.}}]{weltman2020fundamental}%
  \BibitemOpen
  \bibfield  {author} {\bibinfo {author} {\bibfnamefont {A.}~\bibnamefont
  {Weltman}}, \bibinfo {author} {\bibfnamefont {P.}~\bibnamefont {Bull}},
  \bibinfo {author} {\bibfnamefont {S.}~\bibnamefont {Camera}}, \bibinfo
  {author} {\bibfnamefont {K.}~\bibnamefont {Kelley}}, \bibinfo {author}
  {\bibfnamefont {H.}~\bibnamefont {Padmanabhan}}, \bibinfo {author}
  {\bibfnamefont {J.}~\bibnamefont {Pritchard}}, \bibinfo {author}
  {\bibfnamefont {A.}~\bibnamefont {Raccanelli}}, \bibinfo {author}
  {\bibfnamefont {S.}~\bibnamefont {Riemer-S{\o}rensen}}, \bibinfo {author}
  {\bibfnamefont {L.}~\bibnamefont {Shao}}, \bibinfo {author} {\bibfnamefont
  {S.}~\bibnamefont {Andrianomena}}, \emph {et~al.},\ }\bibfield  {title}
  {\bibinfo {title} {Fundamental physics with the square kilometre array},\
  }\href@noop {} {\bibfield  {journal} {\bibinfo  {journal} {Publications of
  the Astronomical Society of Australia}\ }\textbf {\bibinfo {volume} {37}},\
  \bibinfo {pages} {e002} (\bibinfo {year} {2020})}\BibitemShut {NoStop}%
\bibitem [{\citenamefont {Amiaux}\ \emph {et~al.}(2012)\citenamefont {Amiaux},
  \citenamefont {Scaramella}, \citenamefont {Mellier}, \citenamefont {Altieri},
  \citenamefont {Burigana}, \citenamefont {Da~Silva}, \citenamefont {Gomez},
  \citenamefont {Hoar}, \citenamefont {Laureijs}, \citenamefont {Maiorano}
  \emph {et~al.}}]{amiaux2012euclid}%
  \BibitemOpen
  \bibfield  {author} {\bibinfo {author} {\bibfnamefont {J.}~\bibnamefont
  {Amiaux}}, \bibinfo {author} {\bibfnamefont {R.}~\bibnamefont {Scaramella}},
  \bibinfo {author} {\bibfnamefont {Y.}~\bibnamefont {Mellier}}, \bibinfo
  {author} {\bibfnamefont {B.}~\bibnamefont {Altieri}}, \bibinfo {author}
  {\bibfnamefont {C.}~\bibnamefont {Burigana}}, \bibinfo {author}
  {\bibfnamefont {A.}~\bibnamefont {Da~Silva}}, \bibinfo {author}
  {\bibfnamefont {P.}~\bibnamefont {Gomez}}, \bibinfo {author} {\bibfnamefont
  {J.}~\bibnamefont {Hoar}}, \bibinfo {author} {\bibfnamefont {R.}~\bibnamefont
  {Laureijs}}, \bibinfo {author} {\bibfnamefont {E.}~\bibnamefont {Maiorano}},
  \emph {et~al.},\ }\bibfield  {title} {\bibinfo {title} {Euclid mission:
  building of a reference survey},\ }in\ \href@noop {} {\emph {\bibinfo
  {booktitle} {Space Telescopes and Instrumentation 2012: Optical, Infrared,
  and Millimeter Wave}}},\ Vol.\ \bibinfo {volume} {8442}\ (\bibinfo
  {organization} {SPIE},\ \bibinfo {year} {2012})\ pp.\ \bibinfo {pages}
  {380--390}\BibitemShut {NoStop}%
\bibitem [{\citenamefont {Aghamousa}\ \emph {et~al.}(2016)\citenamefont
  {Aghamousa}, \citenamefont {Aguilar}, \citenamefont {Ahlen}, \citenamefont
  {Alam}, \citenamefont {Allen}, \citenamefont {Prieto}, \citenamefont {Annis},
  \citenamefont {Bailey}, \citenamefont {Balland}, \citenamefont {Ballester}
  \emph {et~al.}}]{aghamousa2016desi}%
  \BibitemOpen
  \bibfield  {author} {\bibinfo {author} {\bibfnamefont {A.}~\bibnamefont
  {Aghamousa}}, \bibinfo {author} {\bibfnamefont {J.}~\bibnamefont {Aguilar}},
  \bibinfo {author} {\bibfnamefont {S.}~\bibnamefont {Ahlen}}, \bibinfo
  {author} {\bibfnamefont {S.}~\bibnamefont {Alam}}, \bibinfo {author}
  {\bibfnamefont {L.~E.}\ \bibnamefont {Allen}}, \bibinfo {author}
  {\bibfnamefont {C.~A.}\ \bibnamefont {Prieto}}, \bibinfo {author}
  {\bibfnamefont {J.}~\bibnamefont {Annis}}, \bibinfo {author} {\bibfnamefont
  {S.}~\bibnamefont {Bailey}}, \bibinfo {author} {\bibfnamefont
  {C.}~\bibnamefont {Balland}}, \bibinfo {author} {\bibfnamefont
  {O.}~\bibnamefont {Ballester}}, \emph {et~al.},\ }\bibfield  {title}
  {\bibinfo {title} {The desi experiment part i: science, targeting, and survey
  design},\ }\href@noop {} {\bibfield  {journal} {\bibinfo  {journal}
  {arXiv:1611.00036}\ } (\bibinfo {year} {2016})}\BibitemShut {NoStop}%
\bibitem [{\citenamefont {Abazajian}\ \emph {et~al.}(2019)\citenamefont
  {Abazajian}, \citenamefont {Addison}, \citenamefont {Adshead}, \citenamefont
  {Ahmed}, \citenamefont {Allen}, \citenamefont {Alonso}, \citenamefont
  {Alvarez}, \citenamefont {Anderson}, \citenamefont {Arnold}, \citenamefont
  {Baccigalupi} \emph {et~al.}}]{abazajian2019cmb}%
  \BibitemOpen
  \bibfield  {author} {\bibinfo {author} {\bibfnamefont {K.}~\bibnamefont
  {Abazajian}}, \bibinfo {author} {\bibfnamefont {G.}~\bibnamefont {Addison}},
  \bibinfo {author} {\bibfnamefont {P.}~\bibnamefont {Adshead}}, \bibinfo
  {author} {\bibfnamefont {Z.}~\bibnamefont {Ahmed}}, \bibinfo {author}
  {\bibfnamefont {S.~W.}\ \bibnamefont {Allen}}, \bibinfo {author}
  {\bibfnamefont {D.}~\bibnamefont {Alonso}}, \bibinfo {author} {\bibfnamefont
  {M.}~\bibnamefont {Alvarez}}, \bibinfo {author} {\bibfnamefont
  {A.}~\bibnamefont {Anderson}}, \bibinfo {author} {\bibfnamefont {K.~S.}\
  \bibnamefont {Arnold}}, \bibinfo {author} {\bibfnamefont {C.}~\bibnamefont
  {Baccigalupi}}, \emph {et~al.},\ }\bibfield  {title} {\bibinfo {title}
  {Cmb-s4 science case, reference design, and project plan},\ }\href@noop {}
  {\bibfield  {journal} {\bibinfo  {journal} {arXiv:1907.04473}\ } (\bibinfo
  {year} {2019})}\BibitemShut {NoStop}%
\bibitem [{\citenamefont {Ivezi{\'c}}\ \emph {et~al.}(2019)\citenamefont
  {Ivezi{\'c}}, \citenamefont {Kahn}, \citenamefont {Tyson}, \citenamefont
  {Abel}, \citenamefont {Acosta}, \citenamefont {Allsman}, \citenamefont
  {Alonso}, \citenamefont {AlSayyad}, \citenamefont {Anderson}, \citenamefont
  {Andrew} \emph {et~al.}}]{ivezic2019lsst}%
  \BibitemOpen
  \bibfield  {author} {\bibinfo {author} {\bibfnamefont {{\v{Z}}.}~\bibnamefont
  {Ivezi{\'c}}}, \bibinfo {author} {\bibfnamefont {S.~M.}\ \bibnamefont
  {Kahn}}, \bibinfo {author} {\bibfnamefont {J.~A.}\ \bibnamefont {Tyson}},
  \bibinfo {author} {\bibfnamefont {B.}~\bibnamefont {Abel}}, \bibinfo {author}
  {\bibfnamefont {E.}~\bibnamefont {Acosta}}, \bibinfo {author} {\bibfnamefont
  {R.}~\bibnamefont {Allsman}}, \bibinfo {author} {\bibfnamefont
  {D.}~\bibnamefont {Alonso}}, \bibinfo {author} {\bibfnamefont
  {Y.}~\bibnamefont {AlSayyad}}, \bibinfo {author} {\bibfnamefont {S.~F.}\
  \bibnamefont {Anderson}}, \bibinfo {author} {\bibfnamefont {J.}~\bibnamefont
  {Andrew}}, \emph {et~al.},\ }\bibfield  {title} {\bibinfo {title} {Lsst: from
  science drivers to reference design and anticipated data products},\
  }\href@noop {} {\bibfield  {journal} {\bibinfo  {journal} {The Astrophysical
  Journal}\ }\textbf {\bibinfo {volume} {873}},\ \bibinfo {pages} {111}
  (\bibinfo {year} {2019})}\BibitemShut {NoStop}%
\bibitem [{\citenamefont {Mosby~Jr}\ \emph {et~al.}(2020)\citenamefont
  {Mosby~Jr}, \citenamefont {Rauscher}, \citenamefont {Bennett}, \citenamefont
  {Cheng}, \citenamefont {Cheung}, \citenamefont {Cillis}, \citenamefont
  {Content}, \citenamefont {Cottingham}, \citenamefont {Foltz}, \citenamefont
  {Gygax} \emph {et~al.}}]{mosby2020properties}%
  \BibitemOpen
  \bibfield  {author} {\bibinfo {author} {\bibfnamefont {G.}~\bibnamefont
  {Mosby~Jr}}, \bibinfo {author} {\bibfnamefont {B.~J.}\ \bibnamefont
  {Rauscher}}, \bibinfo {author} {\bibfnamefont {C.}~\bibnamefont {Bennett}},
  \bibinfo {author} {\bibfnamefont {E.~S.}\ \bibnamefont {Cheng}}, \bibinfo
  {author} {\bibfnamefont {S.}~\bibnamefont {Cheung}}, \bibinfo {author}
  {\bibfnamefont {A.}~\bibnamefont {Cillis}}, \bibinfo {author} {\bibfnamefont
  {D.}~\bibnamefont {Content}}, \bibinfo {author} {\bibfnamefont
  {D.}~\bibnamefont {Cottingham}}, \bibinfo {author} {\bibfnamefont
  {R.}~\bibnamefont {Foltz}}, \bibinfo {author} {\bibfnamefont
  {J.}~\bibnamefont {Gygax}}, \emph {et~al.},\ }\bibfield  {title} {\bibinfo
  {title} {Properties and characteristics of the nancy grace roman space
  telescope h4rg-10 detectors},\ }\href@noop {} {\bibfield  {journal} {\bibinfo
   {journal} {Journal of Astronomical Telescopes, Instruments, and Systems}\
  }\textbf {\bibinfo {volume} {6}},\ \bibinfo {pages} {046001} (\bibinfo {year}
  {2020})}\BibitemShut {NoStop}%
\bibitem [{\citenamefont {Newburgh}\ \emph {et~al.}(2016)\citenamefont
  {Newburgh}, \citenamefont {Bandura}, \citenamefont {Bucher}, \citenamefont
  {Chang}, \citenamefont {Chiang}, \citenamefont {Cliche}, \citenamefont
  {Dav{\'e}}, \citenamefont {Dobbs}, \citenamefont {Clarkson}, \citenamefont
  {Ganga} \emph {et~al.}}]{newburgh2016hirax}%
  \BibitemOpen
  \bibfield  {author} {\bibinfo {author} {\bibfnamefont {L.}~\bibnamefont
  {Newburgh}}, \bibinfo {author} {\bibfnamefont {K.}~\bibnamefont {Bandura}},
  \bibinfo {author} {\bibfnamefont {M.}~\bibnamefont {Bucher}}, \bibinfo
  {author} {\bibfnamefont {T.-C.}\ \bibnamefont {Chang}}, \bibinfo {author}
  {\bibfnamefont {H.}~\bibnamefont {Chiang}}, \bibinfo {author} {\bibfnamefont
  {J.}~\bibnamefont {Cliche}}, \bibinfo {author} {\bibfnamefont
  {R.}~\bibnamefont {Dav{\'e}}}, \bibinfo {author} {\bibfnamefont
  {M.}~\bibnamefont {Dobbs}}, \bibinfo {author} {\bibfnamefont
  {C.}~\bibnamefont {Clarkson}}, \bibinfo {author} {\bibfnamefont
  {K.}~\bibnamefont {Ganga}}, \emph {et~al.},\ }\bibfield  {title} {\bibinfo
  {title} {Hirax: a probe of dark energy and radio transients},\ }in\
  \href@noop {} {\emph {\bibinfo {booktitle} {Ground-based and Airborne
  Telescopes VI}}},\ Vol.\ \bibinfo {volume} {9906}\ (\bibinfo {organization}
  {SPIE},\ \bibinfo {year} {2016})\ pp.\ \bibinfo {pages}
  {2039--2049}\BibitemShut {NoStop}%
\bibitem [{\citenamefont {Amiri}\ \emph {et~al.}(2022)\citenamefont {Amiri},
  \citenamefont {Bandura}, \citenamefont {Boskovic}, \citenamefont {Chen},
  \citenamefont {Cliche}, \citenamefont {Deng}, \citenamefont {Denman},
  \citenamefont {Dobbs}, \citenamefont {Fandino}, \citenamefont {Foreman} \emph
  {et~al.}}]{amiri2022overview}%
  \BibitemOpen
  \bibfield  {author} {\bibinfo {author} {\bibfnamefont {M.}~\bibnamefont
  {Amiri}}, \bibinfo {author} {\bibfnamefont {K.}~\bibnamefont {Bandura}},
  \bibinfo {author} {\bibfnamefont {A.}~\bibnamefont {Boskovic}}, \bibinfo
  {author} {\bibfnamefont {T.}~\bibnamefont {Chen}}, \bibinfo {author}
  {\bibfnamefont {J.-F.}\ \bibnamefont {Cliche}}, \bibinfo {author}
  {\bibfnamefont {M.}~\bibnamefont {Deng}}, \bibinfo {author} {\bibfnamefont
  {N.}~\bibnamefont {Denman}}, \bibinfo {author} {\bibfnamefont
  {M.}~\bibnamefont {Dobbs}}, \bibinfo {author} {\bibfnamefont
  {M.}~\bibnamefont {Fandino}}, \bibinfo {author} {\bibfnamefont
  {S.}~\bibnamefont {Foreman}}, \emph {et~al.},\ }\bibfield  {title} {\bibinfo
  {title} {An overview of chime, the canadian hydrogen intensity mapping
  experiment},\ }\href@noop {} {\bibfield  {journal} {\bibinfo  {journal} {The
  Astrophysical Journal Supplement Series}\ }\textbf {\bibinfo {volume}
  {261}},\ \bibinfo {pages} {29} (\bibinfo {year} {2022})}\BibitemShut
  {NoStop}%
\bibitem [{\citenamefont {Dor{\'e}}\ \emph {et~al.}(2014)\citenamefont
  {Dor{\'e}}, \citenamefont {Bock}, \citenamefont {Ashby}, \citenamefont
  {Capak}, \citenamefont {Cooray}, \citenamefont {de~Putter}, \citenamefont
  {Eifler}, \citenamefont {Flagey}, \citenamefont {Gong}, \citenamefont {Habib}
  \emph {et~al.}}]{dore2014cosmology}%
  \BibitemOpen
  \bibfield  {author} {\bibinfo {author} {\bibfnamefont {O.}~\bibnamefont
  {Dor{\'e}}}, \bibinfo {author} {\bibfnamefont {J.}~\bibnamefont {Bock}},
  \bibinfo {author} {\bibfnamefont {M.}~\bibnamefont {Ashby}}, \bibinfo
  {author} {\bibfnamefont {P.}~\bibnamefont {Capak}}, \bibinfo {author}
  {\bibfnamefont {A.}~\bibnamefont {Cooray}}, \bibinfo {author} {\bibfnamefont
  {R.}~\bibnamefont {de~Putter}}, \bibinfo {author} {\bibfnamefont
  {T.}~\bibnamefont {Eifler}}, \bibinfo {author} {\bibfnamefont
  {N.}~\bibnamefont {Flagey}}, \bibinfo {author} {\bibfnamefont
  {Y.}~\bibnamefont {Gong}}, \bibinfo {author} {\bibfnamefont {S.}~\bibnamefont
  {Habib}}, \emph {et~al.},\ }\bibfield  {title} {\bibinfo {title} {Cosmology
  with the spherex all-sky spectral survey},\ }\href@noop {} {\bibfield
  {journal} {\bibinfo  {journal} {arXiv:1412.4872}\ } (\bibinfo {year}
  {2014})}\BibitemShut {NoStop}%
\bibitem [{\citenamefont {{Vogelsberger}}\ \emph {et~al.}(2020)\citenamefont
  {{Vogelsberger}}, \citenamefont {{Marinacci}}, \citenamefont {{Torrey}},\
  and\ \citenamefont {{Puchwein}}}]{Vogelsberger_review}%
  \BibitemOpen
  \bibfield  {author} {\bibinfo {author} {\bibfnamefont {M.}~\bibnamefont
  {{Vogelsberger}}}, \bibinfo {author} {\bibfnamefont {F.}~\bibnamefont
  {{Marinacci}}}, \bibinfo {author} {\bibfnamefont {P.}~\bibnamefont
  {{Torrey}}},\ and\ \bibinfo {author} {\bibfnamefont {E.}~\bibnamefont
  {{Puchwein}}},\ }\bibfield  {title} {\bibinfo {title} {{Cosmological
  simulations of galaxy formation}},\ }\href
  {https://doi.org/10.1038/s42254-019-0127-2} {\bibfield  {journal} {\bibinfo
  {journal} {Nature Reviews Physics}\ }\textbf {\bibinfo {volume} {2}},\
  \bibinfo {pages} {42} (\bibinfo {year} {2020})},\ \Eprint
  {https://arxiv.org/abs/1909.07976} {arXiv:1909.07976 [astro-ph.GA]}
  \BibitemShut {NoStop}%
\bibitem [{\citenamefont {Villaescusa-Navarro}\ \emph
  {et~al.}(2021)\citenamefont {Villaescusa-Navarro}, \citenamefont
  {Angl{\'e}s-Alc{\'a}zar}, \citenamefont {Genel}, \citenamefont {Spergel},
  \citenamefont {Somerville}, \citenamefont {Dave}, \citenamefont {Pillepich},
  \citenamefont {Hernquist}, \citenamefont {Nelson}, \citenamefont {Torrey}
  \emph {et~al.}}]{villaescusa2021camels}%
  \BibitemOpen
  \bibfield  {author} {\bibinfo {author} {\bibfnamefont {F.}~\bibnamefont
  {Villaescusa-Navarro}}, \bibinfo {author} {\bibfnamefont {D.}~\bibnamefont
  {Angl{\'e}s-Alc{\'a}zar}}, \bibinfo {author} {\bibfnamefont {S.}~\bibnamefont
  {Genel}}, \bibinfo {author} {\bibfnamefont {D.~N.}\ \bibnamefont {Spergel}},
  \bibinfo {author} {\bibfnamefont {R.~S.}\ \bibnamefont {Somerville}},
  \bibinfo {author} {\bibfnamefont {R.}~\bibnamefont {Dave}}, \bibinfo {author}
  {\bibfnamefont {A.}~\bibnamefont {Pillepich}}, \bibinfo {author}
  {\bibfnamefont {L.}~\bibnamefont {Hernquist}}, \bibinfo {author}
  {\bibfnamefont {D.}~\bibnamefont {Nelson}}, \bibinfo {author} {\bibfnamefont
  {P.}~\bibnamefont {Torrey}}, \emph {et~al.},\ }\bibfield  {title} {\bibinfo
  {title} {The camels project: Cosmology and astrophysics with machine-learning
  simulations},\ }\href@noop {} {\bibfield  {journal} {\bibinfo  {journal} {The
  Astrophysical Journal}\ }\textbf {\bibinfo {volume} {915}},\ \bibinfo {pages}
  {71} (\bibinfo {year} {2021})}\BibitemShut {NoStop}%
\bibitem [{\citenamefont {Iyer}\ \emph {et~al.}(2020)\citenamefont {Iyer},
  \citenamefont {Tacchella}, \citenamefont {Genel}, \citenamefont {Hayward},
  \citenamefont {Hernquist}, \citenamefont {Brooks}, \citenamefont {Caplar},
  \citenamefont {Dav{\'e}}, \citenamefont {Diemer}, \citenamefont {Forbes}
  \emph {et~al.}}]{iyer2020diversity}%
  \BibitemOpen
  \bibfield  {author} {\bibinfo {author} {\bibfnamefont {K.~G.}\ \bibnamefont
  {Iyer}}, \bibinfo {author} {\bibfnamefont {S.}~\bibnamefont {Tacchella}},
  \bibinfo {author} {\bibfnamefont {S.}~\bibnamefont {Genel}}, \bibinfo
  {author} {\bibfnamefont {C.~C.}\ \bibnamefont {Hayward}}, \bibinfo {author}
  {\bibfnamefont {L.}~\bibnamefont {Hernquist}}, \bibinfo {author}
  {\bibfnamefont {A.~M.}\ \bibnamefont {Brooks}}, \bibinfo {author}
  {\bibfnamefont {N.}~\bibnamefont {Caplar}}, \bibinfo {author} {\bibfnamefont
  {R.}~\bibnamefont {Dav{\'e}}}, \bibinfo {author} {\bibfnamefont
  {B.}~\bibnamefont {Diemer}}, \bibinfo {author} {\bibfnamefont {J.~C.}\
  \bibnamefont {Forbes}}, \emph {et~al.},\ }\bibfield  {title} {\bibinfo
  {title} {The diversity and variability of star formation histories in models
  of galaxy evolution},\ }\href@noop {} {\bibfield  {journal} {\bibinfo
  {journal} {Monthly Notices of the Royal Astronomical Society}\ }\textbf
  {\bibinfo {volume} {498}},\ \bibinfo {pages} {430} (\bibinfo {year}
  {2020})}\BibitemShut {NoStop}%
\bibitem [{\citenamefont {Rodriguez}\ \emph {et~al.}(2018)\citenamefont
  {Rodriguez}, \citenamefont {Kacprzak}, \citenamefont {Lucchi}, \citenamefont
  {Amara}, \citenamefont {Sgier}, \citenamefont {Fluri}, \citenamefont
  {Hofmann},\ and\ \citenamefont {R{\'e}fr{\'e}gier}}]{rodriguez2018fast}%
  \BibitemOpen
  \bibfield  {author} {\bibinfo {author} {\bibfnamefont {A.~C.}\ \bibnamefont
  {Rodriguez}}, \bibinfo {author} {\bibfnamefont {T.}~\bibnamefont {Kacprzak}},
  \bibinfo {author} {\bibfnamefont {A.}~\bibnamefont {Lucchi}}, \bibinfo
  {author} {\bibfnamefont {A.}~\bibnamefont {Amara}}, \bibinfo {author}
  {\bibfnamefont {R.}~\bibnamefont {Sgier}}, \bibinfo {author} {\bibfnamefont
  {J.}~\bibnamefont {Fluri}}, \bibinfo {author} {\bibfnamefont
  {T.}~\bibnamefont {Hofmann}},\ and\ \bibinfo {author} {\bibfnamefont
  {A.}~\bibnamefont {R{\'e}fr{\'e}gier}},\ }\bibfield  {title} {\bibinfo
  {title} {Fast cosmic web simulations with generative adversarial networks},\
  }\href@noop {} {\bibfield  {journal} {\bibinfo  {journal} {Computational
  Astrophysics and Cosmology}\ }\textbf {\bibinfo {volume} {5}},\ \bibinfo
  {pages} {1} (\bibinfo {year} {2018})}\BibitemShut {NoStop}%
\bibitem [{\citenamefont {Mustafa}\ \emph {et~al.}(2019)\citenamefont
  {Mustafa}, \citenamefont {Bard}, \citenamefont {Bhimji}, \citenamefont
  {Luki{\'c}}, \citenamefont {Al-Rfou},\ and\ \citenamefont
  {Kratochvil}}]{mustafa2019cosmogan}%
  \BibitemOpen
  \bibfield  {author} {\bibinfo {author} {\bibfnamefont {M.}~\bibnamefont
  {Mustafa}}, \bibinfo {author} {\bibfnamefont {D.}~\bibnamefont {Bard}},
  \bibinfo {author} {\bibfnamefont {W.}~\bibnamefont {Bhimji}}, \bibinfo
  {author} {\bibfnamefont {Z.}~\bibnamefont {Luki{\'c}}}, \bibinfo {author}
  {\bibfnamefont {R.}~\bibnamefont {Al-Rfou}},\ and\ \bibinfo {author}
  {\bibfnamefont {J.~M.}\ \bibnamefont {Kratochvil}},\ }\bibfield  {title}
  {\bibinfo {title} {Cosmogan: creating high-fidelity weak lensing convergence
  maps using generative adversarial networks},\ }\href@noop {} {\bibfield
  {journal} {\bibinfo  {journal} {Computational Astrophysics and Cosmology}\
  }\textbf {\bibinfo {volume} {6}},\ \bibinfo {pages} {1} (\bibinfo {year}
  {2019})}\BibitemShut {NoStop}%
\bibitem [{\citenamefont {Zamudio-Fernandez}\ \emph {et~al.}(2019)\citenamefont
  {Zamudio-Fernandez}, \citenamefont {Okan}, \citenamefont
  {Villaescusa-Navarro}, \citenamefont {Bilaloglu}, \citenamefont {Cengiz},
  \citenamefont {He}, \citenamefont {Levasseur},\ and\ \citenamefont
  {Ho}}]{zamudio2019higan}%
  \BibitemOpen
  \bibfield  {author} {\bibinfo {author} {\bibfnamefont {J.}~\bibnamefont
  {Zamudio-Fernandez}}, \bibinfo {author} {\bibfnamefont {A.}~\bibnamefont
  {Okan}}, \bibinfo {author} {\bibfnamefont {F.}~\bibnamefont
  {Villaescusa-Navarro}}, \bibinfo {author} {\bibfnamefont {S.}~\bibnamefont
  {Bilaloglu}}, \bibinfo {author} {\bibfnamefont {A.~D.}\ \bibnamefont
  {Cengiz}}, \bibinfo {author} {\bibfnamefont {S.}~\bibnamefont {He}}, \bibinfo
  {author} {\bibfnamefont {L.~P.}\ \bibnamefont {Levasseur}},\ and\ \bibinfo
  {author} {\bibfnamefont {S.}~\bibnamefont {Ho}},\ }\bibfield  {title}
  {\bibinfo {title} {Higan: Cosmic neutral hydrogen with generative adversarial
  networks},\ }\href@noop {} {\bibfield  {journal} {\bibinfo  {journal}
  {arXiv:1904.12846}\ } (\bibinfo {year} {2019})}\BibitemShut {NoStop}%
\bibitem [{\citenamefont {Perraudin}\ \emph {et~al.}(2019)\citenamefont
  {Perraudin}, \citenamefont {Srivastava}, \citenamefont {Lucchi},
  \citenamefont {Kacprzak}, \citenamefont {Hofmann},\ and\ \citenamefont
  {R{\'e}fr{\'e}gier}}]{perraudin2019cosmological}%
  \BibitemOpen
  \bibfield  {author} {\bibinfo {author} {\bibfnamefont {N.}~\bibnamefont
  {Perraudin}}, \bibinfo {author} {\bibfnamefont {A.}~\bibnamefont
  {Srivastava}}, \bibinfo {author} {\bibfnamefont {A.}~\bibnamefont {Lucchi}},
  \bibinfo {author} {\bibfnamefont {T.}~\bibnamefont {Kacprzak}}, \bibinfo
  {author} {\bibfnamefont {T.}~\bibnamefont {Hofmann}},\ and\ \bibinfo {author}
  {\bibfnamefont {A.}~\bibnamefont {R{\'e}fr{\'e}gier}},\ }\bibfield  {title}
  {\bibinfo {title} {Cosmological n-body simulations: a challenge for scalable
  generative models},\ }\href@noop {} {\bibfield  {journal} {\bibinfo
  {journal} {Computational Astrophysics and Cosmology}\ }\textbf {\bibinfo
  {volume} {6}},\ \bibinfo {pages} {1} (\bibinfo {year} {2019})}\BibitemShut
  {NoStop}%
\bibitem [{\citenamefont {Feder}\ \emph {et~al.}(2020)\citenamefont {Feder},
  \citenamefont {Berger},\ and\ \citenamefont {Stein}}]{feder2020nonlinear}%
  \BibitemOpen
  \bibfield  {author} {\bibinfo {author} {\bibfnamefont {R.~M.}\ \bibnamefont
  {Feder}}, \bibinfo {author} {\bibfnamefont {P.}~\bibnamefont {Berger}},\ and\
  \bibinfo {author} {\bibfnamefont {G.}~\bibnamefont {Stein}},\ }\bibfield
  {title} {\bibinfo {title} {Nonlinear 3d cosmic web simulation with
  heavy-tailed generative adversarial networks},\ }\href@noop {} {\bibfield
  {journal} {\bibinfo  {journal} {Physical Review D}\ }\textbf {\bibinfo
  {volume} {102}},\ \bibinfo {pages} {103504} (\bibinfo {year}
  {2020})}\BibitemShut {NoStop}%
\bibitem [{\citenamefont {Springel}(2005)}]{springel2005cosmological}%
  \BibitemOpen
  \bibfield  {author} {\bibinfo {author} {\bibfnamefont {V.}~\bibnamefont
  {Springel}},\ }\bibfield  {title} {\bibinfo {title} {The cosmological
  simulation code gadget-2},\ }\href@noop {} {\bibfield  {journal} {\bibinfo
  {journal} {Monthly notices of the royal astronomical society}\ }\textbf
  {\bibinfo {volume} {364}},\ \bibinfo {pages} {1105} (\bibinfo {year}
  {2005})}\BibitemShut {NoStop}%
\bibitem [{\citenamefont {Perraudin}\ \emph {et~al.}(2021)\citenamefont
  {Perraudin}, \citenamefont {Marcon}, \citenamefont {Lucchi},\ and\
  \citenamefont {Kacprzak}}]{perraudin2021emulation}%
  \BibitemOpen
  \bibfield  {author} {\bibinfo {author} {\bibfnamefont {N.}~\bibnamefont
  {Perraudin}}, \bibinfo {author} {\bibfnamefont {S.}~\bibnamefont {Marcon}},
  \bibinfo {author} {\bibfnamefont {A.}~\bibnamefont {Lucchi}},\ and\ \bibinfo
  {author} {\bibfnamefont {T.}~\bibnamefont {Kacprzak}},\ }\bibfield  {title}
  {\bibinfo {title} {Emulation of cosmological mass maps with conditional
  generative adversarial networks},\ }\href@noop {} {\bibfield  {journal}
  {\bibinfo  {journal} {Frontiers in Artificial Intelligence}\ }\textbf
  {\bibinfo {volume} {4}},\ \bibinfo {pages} {673062} (\bibinfo {year}
  {2021})}\BibitemShut {NoStop}%
\bibitem [{\citenamefont {Curtis}\ \emph {et~al.}(2022)\citenamefont {Curtis},
  \citenamefont {Brainerd},\ and\ \citenamefont
  {Hernandez}}]{curtis2022cosmic}%
  \BibitemOpen
  \bibfield  {author} {\bibinfo {author} {\bibfnamefont {O.}~\bibnamefont
  {Curtis}}, \bibinfo {author} {\bibfnamefont {T.~G.}\ \bibnamefont
  {Brainerd}},\ and\ \bibinfo {author} {\bibfnamefont {A.}~\bibnamefont
  {Hernandez}},\ }\bibfield  {title} {\bibinfo {title} {Cosmic voids in
  gan-generated maps of large-scale structure},\ }\href@noop {} {\bibfield
  {journal} {\bibinfo  {journal} {Astronomy and Computing}\ }\textbf {\bibinfo
  {volume} {38}},\ \bibinfo {pages} {100525} (\bibinfo {year}
  {2022})}\BibitemShut {NoStop}%
\bibitem [{\citenamefont {Tr{\"o}ster}\ \emph {et~al.}(2019)\citenamefont
  {Tr{\"o}ster}, \citenamefont {Ferguson}, \citenamefont {Harnois-D{\'e}raps},\
  and\ \citenamefont {McCarthy}}]{troster2019painting}%
  \BibitemOpen
  \bibfield  {author} {\bibinfo {author} {\bibfnamefont {T.}~\bibnamefont
  {Tr{\"o}ster}}, \bibinfo {author} {\bibfnamefont {C.}~\bibnamefont
  {Ferguson}}, \bibinfo {author} {\bibfnamefont {J.}~\bibnamefont
  {Harnois-D{\'e}raps}},\ and\ \bibinfo {author} {\bibfnamefont {I.~G.}\
  \bibnamefont {McCarthy}},\ }\bibfield  {title} {\bibinfo {title} {Painting
  with baryons: augmenting n-body simulations with gas using deep generative
  models},\ }\href@noop {} {\bibfield  {journal} {\bibinfo  {journal} {Monthly
  Notices of the Royal Astronomical Society: Letters}\ }\textbf {\bibinfo
  {volume} {487}},\ \bibinfo {pages} {L24} (\bibinfo {year}
  {2019})}\BibitemShut {NoStop}%
\bibitem [{\citenamefont {Harrington}\ \emph {et~al.}(2022)\citenamefont
  {Harrington}, \citenamefont {Mustafa}, \citenamefont {Dornfest},
  \citenamefont {Horowitz},\ and\ \citenamefont
  {Luki{\'c}}}]{harrington2022fast}%
  \BibitemOpen
  \bibfield  {author} {\bibinfo {author} {\bibfnamefont {P.}~\bibnamefont
  {Harrington}}, \bibinfo {author} {\bibfnamefont {M.}~\bibnamefont {Mustafa}},
  \bibinfo {author} {\bibfnamefont {M.}~\bibnamefont {Dornfest}}, \bibinfo
  {author} {\bibfnamefont {B.}~\bibnamefont {Horowitz}},\ and\ \bibinfo
  {author} {\bibfnamefont {Z.}~\bibnamefont {Luki{\'c}}},\ }\bibfield  {title}
  {\bibinfo {title} {Fast, high-fidelity ly$\alpha$ forests with convolutional
  neural networks},\ }\href@noop {} {\bibfield  {journal} {\bibinfo  {journal}
  {The Astrophysical Journal}\ }\textbf {\bibinfo {volume} {929}},\ \bibinfo
  {pages} {160} (\bibinfo {year} {2022})}\BibitemShut {NoStop}%
\bibitem [{\citenamefont {Andrianomena}\ \emph {et~al.}(2023)\citenamefont
  {Andrianomena}, \citenamefont {Hassan},\ and\ \citenamefont
  {Villaescusa-Navarro}}]{andrianomena2023invertible}%
  \BibitemOpen
  \bibfield  {author} {\bibinfo {author} {\bibfnamefont {S.}~\bibnamefont
  {Andrianomena}}, \bibinfo {author} {\bibfnamefont {S.}~\bibnamefont
  {Hassan}},\ and\ \bibinfo {author} {\bibfnamefont {F.}~\bibnamefont
  {Villaescusa-Navarro}},\ }\bibfield  {title} {\bibinfo {title} {Invertible
  mapping between fields in camels},\ }\href@noop {} {\bibfield  {journal}
  {\bibinfo  {journal} {arXiv:2303.07473}\ } (\bibinfo {year}
  {2023})}\BibitemShut {NoStop}%
\bibitem [{\citenamefont {Horowitz}\ \emph {et~al.}(2021)\citenamefont
  {Horowitz}, \citenamefont {Dornfest}, \citenamefont {Lukic},\ and\
  \citenamefont {Harrington}}]{horowitz2021hyphy}%
  \BibitemOpen
  \bibfield  {author} {\bibinfo {author} {\bibfnamefont {B.}~\bibnamefont
  {Horowitz}}, \bibinfo {author} {\bibfnamefont {M.}~\bibnamefont {Dornfest}},
  \bibinfo {author} {\bibfnamefont {Z.}~\bibnamefont {Lukic}},\ and\ \bibinfo
  {author} {\bibfnamefont {P.}~\bibnamefont {Harrington}},\ }\bibfield  {title}
  {\bibinfo {title} {Hyphy: Deep generative conditional posterior mapping of
  hydrodynamical physics. arxiv e-prints, art},\ }\href@noop {} {\bibfield
  {journal} {\bibinfo  {journal} {arXiv:2106.12675}\ } (\bibinfo {year}
  {2021})}\BibitemShut {NoStop}%
\bibitem [{\citenamefont {Dai}\ and\ \citenamefont
  {Seljak}(2021)}]{dai2021learning}%
  \BibitemOpen
  \bibfield  {author} {\bibinfo {author} {\bibfnamefont {B.}~\bibnamefont
  {Dai}}\ and\ \bibinfo {author} {\bibfnamefont {U.}~\bibnamefont {Seljak}},\
  }\bibfield  {title} {\bibinfo {title} {Learning effective physical laws for
  generating cosmological hydrodynamics with lagrangian deep learning},\
  }\href@noop {} {\bibfield  {journal} {\bibinfo  {journal} {Proceedings of the
  National Academy of Sciences}\ }\textbf {\bibinfo {volume} {118}},\ \bibinfo
  {pages} {e2020324118} (\bibinfo {year} {2021})}\BibitemShut {NoStop}%
\bibitem [{\citenamefont {Tamosiunas}\ \emph {et~al.}(2021)\citenamefont
  {Tamosiunas}, \citenamefont {Winther}, \citenamefont {Koyama}, \citenamefont
  {Bacon}, \citenamefont {Nichol},\ and\ \citenamefont
  {Mawdsley}}]{tamosiunas2021investigating}%
  \BibitemOpen
  \bibfield  {author} {\bibinfo {author} {\bibfnamefont {A.}~\bibnamefont
  {Tamosiunas}}, \bibinfo {author} {\bibfnamefont {H.~A.}\ \bibnamefont
  {Winther}}, \bibinfo {author} {\bibfnamefont {K.}~\bibnamefont {Koyama}},
  \bibinfo {author} {\bibfnamefont {D.~J.}\ \bibnamefont {Bacon}}, \bibinfo
  {author} {\bibfnamefont {R.~C.}\ \bibnamefont {Nichol}},\ and\ \bibinfo
  {author} {\bibfnamefont {B.}~\bibnamefont {Mawdsley}},\ }\bibfield  {title}
  {\bibinfo {title} {Investigating cosmological gan emulators using latent
  space interpolation},\ }\href@noop {} {\bibfield  {journal} {\bibinfo
  {journal} {Monthly Notices of the Royal Astronomical Society}\ }\textbf
  {\bibinfo {volume} {506}},\ \bibinfo {pages} {3049} (\bibinfo {year}
  {2021})}\BibitemShut {NoStop}%
\bibitem [{\citenamefont {Goodfellow}\ \emph {et~al.}(2014)\citenamefont
  {Goodfellow}, \citenamefont {Pouget-Abadie}, \citenamefont {Mirza},
  \citenamefont {Xu}, \citenamefont {Warde-Farley}, \citenamefont {Ozair},
  \citenamefont {Courville},\ and\ \citenamefont
  {Bengio}}]{goodfellow2014generative}%
  \BibitemOpen
  \bibfield  {author} {\bibinfo {author} {\bibfnamefont {I.}~\bibnamefont
  {Goodfellow}}, \bibinfo {author} {\bibfnamefont {J.}~\bibnamefont
  {Pouget-Abadie}}, \bibinfo {author} {\bibfnamefont {M.}~\bibnamefont
  {Mirza}}, \bibinfo {author} {\bibfnamefont {B.}~\bibnamefont {Xu}}, \bibinfo
  {author} {\bibfnamefont {D.}~\bibnamefont {Warde-Farley}}, \bibinfo {author}
  {\bibfnamefont {S.}~\bibnamefont {Ozair}}, \bibinfo {author} {\bibfnamefont
  {A.}~\bibnamefont {Courville}},\ and\ \bibinfo {author} {\bibfnamefont
  {Y.}~\bibnamefont {Bengio}},\ }\bibfield  {title} {\bibinfo {title}
  {Generative adversarial nets},\ }in\ \href@noop {} {\emph {\bibinfo
  {booktitle} {Advances in neural information processing systems}}}\ (\bibinfo
  {year} {2014})\ pp.\ \bibinfo {pages} {2672--2680}\BibitemShut {NoStop}%
\bibitem [{\citenamefont {Radford}\ \emph {et~al.}(2015)\citenamefont
  {Radford}, \citenamefont {Metz},\ and\ \citenamefont
  {Chintala}}]{radford2015unsupervised}%
  \BibitemOpen
  \bibfield  {author} {\bibinfo {author} {\bibfnamefont {A.}~\bibnamefont
  {Radford}}, \bibinfo {author} {\bibfnamefont {L.}~\bibnamefont {Metz}},\ and\
  \bibinfo {author} {\bibfnamefont {S.}~\bibnamefont {Chintala}},\ }\bibfield
  {title} {\bibinfo {title} {Unsupervised representation learning with deep
  convolutional generative adversarial networks},\ }\href@noop {} {\bibfield
  {journal} {\bibinfo  {journal} {arXiv preprint arXiv:1511.06434}\ } (\bibinfo
  {year} {2015})}\BibitemShut {NoStop}%
\bibitem [{\citenamefont {Liu}\ \emph {et~al.}(2020)\citenamefont {Liu},
  \citenamefont {Zhu}, \citenamefont {Song},\ and\ \citenamefont
  {Elgammal}}]{liu2020towards}%
  \BibitemOpen
  \bibfield  {author} {\bibinfo {author} {\bibfnamefont {B.}~\bibnamefont
  {Liu}}, \bibinfo {author} {\bibfnamefont {Y.}~\bibnamefont {Zhu}}, \bibinfo
  {author} {\bibfnamefont {K.}~\bibnamefont {Song}},\ and\ \bibinfo {author}
  {\bibfnamefont {A.}~\bibnamefont {Elgammal}},\ }\bibfield  {title} {\bibinfo
  {title} {Towards faster and stabilized gan training for high-fidelity
  few-shot image synthesis},\ }in\ \href@noop {} {\emph {\bibinfo {booktitle}
  {International Conference on Learning Representations}}}\ (\bibinfo {year}
  {2020})\BibitemShut {NoStop}%
\bibitem [{\citenamefont {He}\ \emph {et~al.}(2016)\citenamefont {He},
  \citenamefont {Zhang}, \citenamefont {Ren},\ and\ \citenamefont
  {Sun}}]{he2016deep}%
  \BibitemOpen
  \bibfield  {author} {\bibinfo {author} {\bibfnamefont {K.}~\bibnamefont
  {He}}, \bibinfo {author} {\bibfnamefont {X.}~\bibnamefont {Zhang}}, \bibinfo
  {author} {\bibfnamefont {S.}~\bibnamefont {Ren}},\ and\ \bibinfo {author}
  {\bibfnamefont {J.}~\bibnamefont {Sun}},\ }\bibfield  {title} {\bibinfo
  {title} {Deep residual learning for image recognition},\ }in\ \href@noop {}
  {\emph {\bibinfo {booktitle} {Proceedings of the IEEE conference on computer
  vision and pattern recognition}}}\ (\bibinfo {year} {2016})\ pp.\ \bibinfo
  {pages} {770--778}\BibitemShut {NoStop}%
\bibitem [{\citenamefont {Villaescusa-Navarro}\ \emph
  {et~al.}(2022{\natexlab{a}})\citenamefont {Villaescusa-Navarro},
  \citenamefont {Genel}, \citenamefont {Angles-Alcazar}, \citenamefont
  {Thiele}, \citenamefont {Dave}, \citenamefont {Narayanan}, \citenamefont
  {Nicola}, \citenamefont {Li}, \citenamefont {Villanueva-Domingo},
  \citenamefont {Wandelt} \emph {et~al.}}]{CMD}%
  \BibitemOpen
  \bibfield  {author} {\bibinfo {author} {\bibfnamefont {F.}~\bibnamefont
  {Villaescusa-Navarro}}, \bibinfo {author} {\bibfnamefont {S.}~\bibnamefont
  {Genel}}, \bibinfo {author} {\bibfnamefont {D.}~\bibnamefont
  {Angles-Alcazar}}, \bibinfo {author} {\bibfnamefont {L.}~\bibnamefont
  {Thiele}}, \bibinfo {author} {\bibfnamefont {R.}~\bibnamefont {Dave}},
  \bibinfo {author} {\bibfnamefont {D.}~\bibnamefont {Narayanan}}, \bibinfo
  {author} {\bibfnamefont {A.}~\bibnamefont {Nicola}}, \bibinfo {author}
  {\bibfnamefont {Y.}~\bibnamefont {Li}}, \bibinfo {author} {\bibfnamefont
  {P.}~\bibnamefont {Villanueva-Domingo}}, \bibinfo {author} {\bibfnamefont
  {B.}~\bibnamefont {Wandelt}}, \emph {et~al.},\ }\bibfield  {title} {\bibinfo
  {title} {The camels multifield data set: Learning the universe’s
  fundamental parameters with artificial intelligence},\ }\href@noop {}
  {\bibfield  {journal} {\bibinfo  {journal} {The Astrophysical Journal
  Supplement Series}\ }\textbf {\bibinfo {volume} {259}},\ \bibinfo {pages}
  {61} (\bibinfo {year} {2022}{\natexlab{a}})}\BibitemShut {NoStop}%
\bibitem [{\citenamefont {Villaescusa-Navarro}\ \emph
  {et~al.}(2022{\natexlab{b}})\citenamefont {Villaescusa-Navarro},
  \citenamefont {Genel}, \citenamefont {Angl{\'e}s-Alc{\'a}zar}, \citenamefont
  {Perez}, \citenamefont {Villanueva-Domingo}, \citenamefont {Wadekar},
  \citenamefont {Shao}, \citenamefont {Mohammad}, \citenamefont {Hassan},
  \citenamefont {Moser} \emph {et~al.}}]{pacodatarelease}%
  \BibitemOpen
  \bibfield  {author} {\bibinfo {author} {\bibfnamefont {F.}~\bibnamefont
  {Villaescusa-Navarro}}, \bibinfo {author} {\bibfnamefont {S.}~\bibnamefont
  {Genel}}, \bibinfo {author} {\bibfnamefont {D.}~\bibnamefont
  {Angl{\'e}s-Alc{\'a}zar}}, \bibinfo {author} {\bibfnamefont {L.~A.}\
  \bibnamefont {Perez}}, \bibinfo {author} {\bibfnamefont {P.}~\bibnamefont
  {Villanueva-Domingo}}, \bibinfo {author} {\bibfnamefont {D.}~\bibnamefont
  {Wadekar}}, \bibinfo {author} {\bibfnamefont {H.}~\bibnamefont {Shao}},
  \bibinfo {author} {\bibfnamefont {F.~G.}\ \bibnamefont {Mohammad}}, \bibinfo
  {author} {\bibfnamefont {S.}~\bibnamefont {Hassan}}, \bibinfo {author}
  {\bibfnamefont {E.}~\bibnamefont {Moser}}, \emph {et~al.},\ }\bibfield
  {title} {\bibinfo {title} {The camels project: public data release},\
  }\href@noop {} {\bibfield  {journal} {\bibinfo  {journal} {arXiv:2201.01300}\
  } (\bibinfo {year} {2022}{\natexlab{b}})}\BibitemShut {NoStop}%
\bibitem [{\citenamefont {Villaescusa-Navarro}(2018)}]{villaescusa2018pylians}%
  \BibitemOpen
  \bibfield  {author} {\bibinfo {author} {\bibfnamefont {F.}~\bibnamefont
  {Villaescusa-Navarro}},\ }\bibfield  {title} {\bibinfo {title} {Pylians:
  Python libraries for the analysis of numerical simulations},\ }\href@noop {}
  {\bibfield  {journal} {\bibinfo  {journal} {Astrophysics Source Code
  Library}\ ,\ \bibinfo {pages} {ascl}} (\bibinfo {year} {2018})}\BibitemShut
  {NoStop}%
\bibitem [{\citenamefont {Hassan}\ \emph {et~al.}(2021)\citenamefont {Hassan},
  \citenamefont {Villaescusa-Navarro}, \citenamefont {Wandelt}, \citenamefont
  {Spergel}, \citenamefont {Angl{\'e}s-Alc{\'a}zar}, \citenamefont {Genel},
  \citenamefont {Cranmer}, \citenamefont {Bryan}, \citenamefont {Dav{\'e}},
  \citenamefont {Somerville} \emph {et~al.}}]{hassan2021hiflow}%
  \BibitemOpen
  \bibfield  {author} {\bibinfo {author} {\bibfnamefont {S.}~\bibnamefont
  {Hassan}}, \bibinfo {author} {\bibfnamefont {F.}~\bibnamefont
  {Villaescusa-Navarro}}, \bibinfo {author} {\bibfnamefont {B.}~\bibnamefont
  {Wandelt}}, \bibinfo {author} {\bibfnamefont {D.~N.}\ \bibnamefont
  {Spergel}}, \bibinfo {author} {\bibfnamefont {D.}~\bibnamefont
  {Angl{\'e}s-Alc{\'a}zar}}, \bibinfo {author} {\bibfnamefont {S.}~\bibnamefont
  {Genel}}, \bibinfo {author} {\bibfnamefont {M.}~\bibnamefont {Cranmer}},
  \bibinfo {author} {\bibfnamefont {G.~L.}\ \bibnamefont {Bryan}}, \bibinfo
  {author} {\bibfnamefont {R.}~\bibnamefont {Dav{\'e}}}, \bibinfo {author}
  {\bibfnamefont {R.~S.}\ \bibnamefont {Somerville}}, \emph {et~al.},\
  }\bibfield  {title} {\bibinfo {title} {Hiflow: Generating diverse hi maps
  conditioned on cosmology using normalizing flow},\ }\href@noop {} {\bibfield
  {journal} {\bibinfo  {journal} {arXiv:2110.02983}\ } (\bibinfo {year}
  {2021})}\BibitemShut {NoStop}%
\bibitem [{\citenamefont {Andrianomena}\ and\ \citenamefont
  {Hassan}(2023)}]{andrianomena2022predictive}%
  \BibitemOpen
  \bibfield  {author} {\bibinfo {author} {\bibfnamefont {S.}~\bibnamefont
  {Andrianomena}}\ and\ \bibinfo {author} {\bibfnamefont {S.}~\bibnamefont
  {Hassan}},\ }\bibfield  {title} {\bibinfo {title} {Predictive uncertainty on
  astrophysics recovery from multifield cosmology},\ }\href@noop {} {\bibfield
  {journal} {\bibinfo  {journal} {Journal of Cosmology and Astroparticle
  Physics}\ }\textbf {\bibinfo {volume} {2023}}\bibinfo  {number} { (06)},\
  \bibinfo {pages} {051}}\BibitemShut {NoStop}%
\bibitem [{\citenamefont {Hassan}\ \emph {et~al.}(2019)\citenamefont {Hassan},
  \citenamefont {Liu}, \citenamefont {Kohn},\ and\ \citenamefont
  {La~Plante}}]{hassan2019identifying}%
  \BibitemOpen
\bibfield  {number} {  }\bibfield  {author} {\bibinfo {author} {\bibfnamefont
  {S.}~\bibnamefont {Hassan}}, \bibinfo {author} {\bibfnamefont
  {A.}~\bibnamefont {Liu}}, \bibinfo {author} {\bibfnamefont {S.}~\bibnamefont
  {Kohn}},\ and\ \bibinfo {author} {\bibfnamefont {P.}~\bibnamefont
  {La~Plante}},\ }\bibfield  {title} {\bibinfo {title} {Identifying
  reionization sources from 21 cm maps using convolutional neural networks},\
  }\href@noop {} {\bibfield  {journal} {\bibinfo  {journal} {Monthly Notices of
  the Royal Astronomical Society}\ }\textbf {\bibinfo {volume} {483}},\
  \bibinfo {pages} {2524} (\bibinfo {year} {2019})}\BibitemShut {NoStop}%
\bibitem [{\citenamefont {Gondhalekar}\ \emph {et~al.}(2023)\citenamefont
  {Gondhalekar}, \citenamefont {Hassan}, \citenamefont {Saphra},\ and\
  \citenamefont {Andrianomena}}]{gondhalekar2023towards}%
  \BibitemOpen
  \bibfield  {author} {\bibinfo {author} {\bibfnamefont {Y.}~\bibnamefont
  {Gondhalekar}}, \bibinfo {author} {\bibfnamefont {S.}~\bibnamefont {Hassan}},
  \bibinfo {author} {\bibfnamefont {N.}~\bibnamefont {Saphra}},\ and\ \bibinfo
  {author} {\bibfnamefont {S.}~\bibnamefont {Andrianomena}},\ }\bibfield
  {title} {\bibinfo {title} {Towards out-of-distribution generalization in
  large-scale astronomical surveys: robust networks learn similar
  representations},\ }\href@noop {} {\bibfield  {journal} {\bibinfo  {journal}
  {arXiv preprint arXiv:2311.18007}\ } (\bibinfo {year} {2023})}\BibitemShut
  {NoStop}%
\bibitem [{\citenamefont {Ostriker}\ and\ \citenamefont
  {Kim}(2022)}]{ostriker2022pressure}%
  \BibitemOpen
  \bibfield  {author} {\bibinfo {author} {\bibfnamefont {E.~C.}\ \bibnamefont
  {Ostriker}}\ and\ \bibinfo {author} {\bibfnamefont {C.-G.}\ \bibnamefont
  {Kim}},\ }\bibfield  {title} {\bibinfo {title} {Pressure-regulated,
  feedback-modulated star formation in disk galaxies},\ }\href@noop {}
  {\bibfield  {journal} {\bibinfo  {journal} {The Astrophysical Journal}\
  }\textbf {\bibinfo {volume} {936}},\ \bibinfo {pages} {137} (\bibinfo {year}
  {2022})}\BibitemShut {NoStop}%
\end{thebibliography}%

\end{document}